\documentclass{aa}
\usepackage{graphicx}
\usepackage{txfonts}
\usepackage{natbib}
\bibpunct{(}{)}{;}{a}{}{,} 
\begin{document}
\title{OASIS integral-field spectroscopy of the central kpc in 11 
  Seyfert\,2 galaxies}

\author{Ivana Stoklasov\'a \inst{1,2},
  Pierre Ferruit\inst{3},
  Eric Emsellem\inst{3},
  Bruno Jungwiert\inst{1},
  Emmanuel P\'econtal\inst{3},
  Sebastian F. S\'anchez\inst{4}
  }

\offprints{I. Stoklasov\'{a} \\
      \email{ivana@sirrah.troja.mff.cuni.cz}
     }

\institute{Astronomical Institute of the Academy of Sciences 
of the Czech Republic, v.v.i.,
      Bo\v cn\'\i\ II 1401, CZ-14131 Prague, Czech Republic
 \and
     Faculty of Mathematics and Physics, Charles University in Prague,
     Ke Karlovu 3, CZ-12161 Prague, Czech Republic
 \and
    Universit\'e de Lyon, France; Universit\'e Lyon 1, F-69007; CRAL,
    Observatoire de Lyon, F-69230 Saint-Genis-Laval; \\ 
    CNRS, UMR 5574; ENS de Lyon, France         
 \and
     Centro Astron\'omico Hispano Alem\'an, Calar Alto, CSIC-MPG,
     C/Jes\'us Durb\'an Rem\'on 2-2, 04004 Almer{\'\i}a, Spain
     }

\date{Received 27 October 2008 / Accepted 15 January 2009}

\authorrunning{Stoklasov\'a et al.}


\abstract
{}
{We examine the physical conditions of ionised gas in the 
central kiloparsec(s) of
nearby Seyfert galaxies. 
Our aim is a detailed 
two-dimensional mapping of optical spectroscopic properties, 
searching for features in the emission structure and in gas kinematics  
which would be common to all the active galactic nuclei (AGN) or their 
class, and which would help us to understand the origin and 
nature of the circumnuclear gas, and its role in the AGN.      
}
{We map narrow-line regions (NLRs) of 11 nearby Seyfert\,2
galaxies with the optical integral-field spectrograph OASIS 
mounted at CFHT. We model emission-line profiles of 
5 forbidden-line doublets and 2 Balmer lines, correcting
for the underlying stellar absorption by reconstructing  
stellar spectra with synthetic evolutionary stellar population models. 
}
{For each of the 11 targets, we present     
2D maps of surface brightness in the observed emission lines, 
diagnostic line intensity ratios, 
gas kinematics (mean line-of-sight velocity and velocity dispersion),
electron density, and interstellar reddening, and we plot 
spatially resolved spectral-diagnostic diagrams. 
The stellar data are represented
by maps of mean line-of-sight (LOS) velocities and of the relative mass 
fractions of the young stellar populations.   
The gas velocity fields in $80\%$ of the sample exhibit twisted S-shaped 
isovelocity contours, which are signatures of non-circular orbits and 
indicate non-axisymmetric gravitational potentials, gas motions out of
the galactic plane, or possible outflows and inflows. 
Based on the kinematic measurements, we identified a 
possible nuclear ring or radial gas flow in 
NGC 262 (Mrk 348), not reported before. 
Eight of the eleven observed objects exhibit strongly asymmetric 
or multi-component emission-line profiles, in most cases 
confined to an elongated region passing through the galactic 
centre, perpendicular to the major axis of emission.       
}
{}

\keywords{galaxies: active; galaxies: Seyfert; galaxies: kinematics and
dynamics; methods: observational; methods: data analysis; techniques:
spectroscopic}
\maketitle

\section{Introduction}

In the search for the fuelling mechanism of active galactic nuclei (AGN) 
and its trigger, no significant
differences have been found between the host galaxies of AGN and their
quiescent counterparts. Among others, no excess of galactic bars has been
confirmed
in active galaxies \citep{Martini03,Laurikainen04}.
On the other hand, recent studies of galactic centres have reported an
excess of infrared isophotal twists \citep{Hunt} and of twisted isovelocity
contours in gas \citep{Dumas}, which may imply presence of 
non-axisymmetric potentials \citep{Schinnerer00}.
Focus on gas kinematics in the nuclear regions thus seems to be the key to 
understanding the interplay between the host galaxy and the active nucleus, 
even though interpretation of the phenomena on very different spatial
scales needs to be done with extreme care. 
Detailed spatially resolved spectroscopic observations
of the nuclear regions in large samples of galaxies, 
followed by proper modelling are hence necessary. 

The Narrow-Line Regions (NLR) represent low-density gas ionised
by the central AGN \citep[e.g.,][]{Morse96,Evans99} 
and extending beyond the central dusty torus.
The classical NLRs lie within $10-100$\,pc scales and their kinematics
is often perturbed by energy injection from the AGN while gas motions
in the so-called extended NLRs (ENLRs), reaching to kpc and 10\,kpc scales,
are thought to be predominantly gravitational.
The ENLRs are convenient observational targets for spatially resolved
imaging and spectroscopy, especially in nearby Seyfert galaxies.
They play a crucial role in attempts to understand the AGN structure
and evolution, and in testing of the Unified Models of 
AGN \citep{Antonucci93,Urry}.

By completing narrow-band imaging and aperture spectroscopy observations
of large samples of Seyfert galaxies, robust statistical analyses 
of NLR properties have been achieved
\citep{MulObs,HoIII,Schm03Obs}. Similar surveys
with modern spatially resolved spectroscopic techniques are necessary 
to obtain more complex information and achieve a qualitatively 
new level of conclusions.       
As an example, we mention the controversy in optical imaging studies 
by \citet{MulObs}; and \citet{Schm03Obs}, who tested predictions 
of the Unified Model with respect to the NLR size and morphology, and 
obtained results that cast doubts on the simple model of ionisation of 
ambient gas by a cone of radiation.      
However, both studies are affected by the limitations typical of pure
imaging: these include the survey sensitivity or contamination by other 
emission sources, the most important being
gas ionised by young, massive stars that have been found in
centres of many Seyfert galaxies \citep[e.g.,][]{Schm99,Cid04,Boisson04}.

Spatially resolved spectroscopy overcomes many of these difficulties:
ratios of forbidden and permitted emission-line intensities at different   
locations help us to discriminate 
between the ionisation sources and to establish a 
boundary of the AGN-ionised NLR, using ionisation models
\citep[see e.g.,][for a discussion and application]{BennertSey2}. 
Also, spatially resolved kinematics derived from 
the emission lines provides an important insight into the three-dimensional 
structure and dynamics of the central emitting regions.   
The wealth of information contained in such spectroscopic data thus  
allows us to address 
a number of questions related to 
the NLRs, which are important for reconstructing the global picture of AGN. 
An efficient way of acquiring simultaneously the spatially resolved 
optical or infrared (IR) spectroscopic data is provided by the 
integral-field spectroscopy (IFS). 
Integral-field units (IFUs) are becoming 
a standard type of equipment at large telescopes, available with a 
wide variety of technical parameters, such as spectral range, 
spectral and spatial resolution, and field of view.

\bigskip

We present our mapping of the central $\sim$\,kiloparsec(s) of 
11 nearby Seyfert\,2 galaxies observed with the optical integral-field
spectrograph OASIS mounted at the 3.6\,m Canada--France--Hawaii Telescope 
(CFHT). Now mounted at William Herschel Telescope (WHT), OASIS is an 
IFU well-suited to high-resolution observations over a 
restricted field of view (FOV), for a wide range of spectral modes. 
Its use for detailed spectroscopic mapping of the NLRs and the central parts 
of the ENLRs is thus advantageous for understanding the complex
spatial variations in the physical properties of the circumnuclear ionised gas, 
and the implications for the AGN model.

For the observed emission regions, we plot maps of derived 
quantities such as the surface brightness 
in two Balmer lines and five doublets of forbidden emission lines, 
the computed line spectral-diagnostic ratios, 
the kinematic properties, the electron densities, and 
the interstellar reddening. We plot
spatially resolved spectral-diagnostic diagrams for each object 
where possible. 
We provide details of the individual Seyfert\,2 nuclei in 
Sect.\,\ref{results},
confronting our observations with results found in the literature. 
The complete sample is discussed in Sect.\,\ref{discuss}, where we draw
conclusions about various structural and kinematic results.

\section{Observations and data reduction}

\subsection{Observations \label{Obs}} 
The selected 11 targets (Table\,\ref{physpar}) 
correspond to well-known Seyfert 
galaxies with previously published photometric and/or 
spectroscopic data. Their redshifts are in the range 
$0.006<z<0.051\,,$ corresponding to distances of 17\,Mpc -- 220\,Mpc. 
Observations were performed in the years $2000-2002$, with 
the lenslet-array OASIS spectrograph constructed at the Centre de Recherche
Astrophysique de Lyon (CRAL), France, mounted at CFHT 
(Mauna Kea) in the F/8 Cassegrain focus.  
Two spectral domains were explored, with either high ($0.27\arcsec$)
or low ($0.41\arcsec$) spatial sampling, and corresponding fields of view 
of $10\arcsec\times8\arcsec$ and $15\arcsec\times12\arcsec$, respectively. 
Each of the modes provided approximately a thousand spectra per object 
per spectral
domain. The two spectral modes covered the wavelength ranges of 
$4760\,\mathrm{\r{A}}-5558\,\mathrm{\r{A}}$
with $2.15\,\mathrm{\r{A}}$ dispersion (``MR1'' spectral mode), 
and $6210\,\mathrm{\r{A}}-7008\,\mathrm{\r{A}}$ 
with $2.17\,\mathrm{\r{A}}$ dispersion (``MR2'' spectral mode).

Table\,\ref{list} specifies the spectral and spatial configurations used for 
each member of the entire observed set.
An overview of all the configurations of OASIS is available at the dedicated 
web pages of the instrument and at the CFHT pages. Details of the instrumental 
concept are to be found in \citet{Bacon95}, which describes the TIGER 
spectrograph, a direct predecessor of OASIS.

\subsection{Data reduction}
The data were reduced with the dedicated XOasis software
\citep{Pec3D,PecLCL}.
The standard reduction procedure includes bias and dark subtraction (CCD), 
extraction of the spectra using a fitted mask model, wavelength calibration, 
low-frequency flat-fielding, spectro-spatial flat-fielding, 
cosmic-ray removal, homogenisation of the spectral resolution over the field,
sky subtraction, and flux calibration using observations of the photometric
standard stars. Multiple exposures were merged and mosaiced, 
truncating the wavelength domain to a common range in the different fields, 
and combining the spectra (both signal and noise) with optimal weights and
renormalisation.
The data cubes were resampled to a common spatial scale on a rectangular
grid (from the original array of hexagonal lenses) with either 
$0.25\arcsec$ or $0.4\arcsec$
spacing depending on the adopted spatial sampling.  
Orientation was defined to be north up, east to the left.

\begin{table*}[ht] 
\begin{center}
\footnotesize   
\caption{List of observed Seyfert galaxies -- basic data
\label{physpar}
}
\begin{tabular}{lllllllccc} 
\hline
\hline
\noalign{\smallskip}
Object    & Alternative & RA  & DEC 
& Hubble &Redshift $z$ & Velocity $cz$ & $D_L$ & Ang.scale 
\\
& name        & [h m s] &  [d m s]  
& type  &  
& [km\,s$^{-1}$]     &  [Mpc]  & [pc/$\arcsec$]
\\
\noalign{\smallskip}
\hline
\hline
\noalign{\smallskip}
Mrk 34    & MCG +10-15-104 & 10 34 08.6 & $+60$ 01 52  & 
Spiral & $0.0505$ 
& $15\,140\pm90$ & 218 & 956 
\\
Mrk 622   & UGC 04229      & 08 07 41.0 & $+39$ 00 15  &
S0 & $0.0232$ 
& $6964\pm11$ & 99.6 & 461 
\\
Mrk 1066  & UGC 02456      & 02 59 58.6 & $+36$ 49 14  &   
(R)SB(s)0+& $0.0120$ 
& $3605\pm22$ & 47.3 & 224
\\
NGC 262  & Mrk 348        & 00 48 47.1 & $+31$ 57 25  &
SA(s)0/a & $0.0150$
& $4507\pm4$ & 58.2 & 274 
\\
NGC 449  & Mrk 1           & 01 16 07.2 & $+33$ 05 22  &
(R')S? & $0.0159$ 
& $4780\pm2$ & 62.3 & 293
\\
NGC 2273  & Mrk 620         & 06 50 08.6 & $+60$ 50 45  &
SB(r)a:& $0.0061$ 
& $1840\pm4$ & 25.8 & 124 
\\
NGC 2992  & MCG -02-25-014  & 09 45 42.0 & $-14$ 19 35  & 
Sa pec& $0.0077$ 
& $2311\pm4$ & 36.6 & 175 
\\
NGC 3081  & IC 2529         & 09 59 29.5 & $-22$ 49 35 & 
(R$_1$)SAB(r)0/a& $0.0080$ 
& $2391\pm3$ & 37.7 & 180
\\
NGC 4388  & UGC 07520       & 12 25 46.7 & $+12$ 39 44 &
SA(s)b: sp& $0.0084$ 
& $2524\pm1$ &
16.7$^{(*)}$&81$^{(*)}$
\\   
NGC 5728  & MCG -03-37-005  & 14 42 23.9 & $-17$ 15 11 & 
(R$_1$)SAB(r)a& $0.0094$ 
& $2804\pm20$ & 41.9 & 199 
\\
NGC 5929  & UGC 09851       & 15 26 06.1 & $+41$ 40 14 &
Sab: pec& $0.0083$ 
& $2492\pm8$ & 35.7 & 170
\\
\noalign{\smallskip}
\hline
\hline
\end{tabular}
\end{center}
Adopted from the NASA/IPAC Extragalactic Database (NED). The cosmological 
scales assume $\mathrm{H}_0 = 73\,\mathrm{km\,s}^{-1}\,\mathrm{Mpc}^{-1}$,
$\Omega_M = 0.27$, $\Omega_\Lambda = 0.73$\,. $D_L$: luminosity distance;
Ang.scale: angular scale.
(*) Values from \citet{Yasuda}, with NGC 4388 considered inside the Virgo
cluster. 
\\
\end{table*} 


\begin{table*}[ht] 
\footnotesize   
\caption{List of observed Seyfert galaxies -- observational parameters.
\label{list}
}
\begin{center}
\begin{tabular}{llllcc} 
\hline
\hline
\noalign{\smallskip}
Object & 
Observation & Integration time &  Integration time &
Spatial sampling & Spatial sampling
\\
& 
date     &  in MR1 [s]  &  in MR2 [s]  &
MR1 & MR2
\\
\noalign{\smallskip}
\hline
\hline
\noalign{\smallskip}
Mrk 34 & 
14-15/03/2001    &  $1\times 2700$  & $2\times 2700$ &
HR & LR
\\
Mrk 622& 
24/11/2000       &  $2\times 1800$  & $2\times 1800$ &
HR & HR
\\
Mrk 1066& 
22/11/2000       &  $3\times 1800 + 1\times 1200$ & $2\times 1800$ &
HR & HR
\\
NGC 262 & 
25/11/2000       &  $2\times 1800$  & $2\times 1800$ &
HR & HR
\\
NGC 449 & 
24/11/2000      & 3 fields $\times\,2  \times 1800$ &
       3 fields $\times\,2  \times 1800$ &
HR & HR 
\\
NGC 2273& 
24/11/2000       &  $2\times 1800$ & --- &
LR & ---
\\
NGC 2992& 
16-17/03/2001    &  $1\times 1800 + 1\times 2700$ & $2\times 1800$ &
LR & HR
\\
NGC 3081& 
17/01/2002       &  $2\times3600$     & ---  &
LR & --- 
\\
NGC 4388& 
17/01/2002       &  $3\times3300$     & --- &
LR & ---
\\   
NGC 5728& 
15/03/2001       &  $2\times 1800$  & $1\times 1800 + 1\times 2700 $ &
LR & LR
\\
NGC 5929& 
16/03/2001 &  $2\times 2700$  & --- &
LR & ---
\\
\noalign{\smallskip}
\hline
\hline
\end{tabular}
\end{center}
The abbreviations used: MR1: spectral mode of wavelength range
$(4760,5558)$\,\r{A}, dispersion $2.15$\,\r{A}/pixel, resolving power
$1210$; MR2: spectral mode of wavelength range $(6210,7008)$\,\r{A},
dispersion
$2.17$\,\r{A}/pixel, resolving power $1525$; 
HR: high spatial resolution mode
with sky sampling $0.27\arcsec$, field of view 
$10.4\arcsec\times8.3\arcsec$; LR: low
spatial resolution mode, sky sampling $0.41\arcsec$, field of view 
$15\arcsec \times 12\arcsec$.
\end{table*}

\section{Data analysis}

The results presented in this paper were obtained by a two-stage analysis
procedure: 
stellar-population modelling and emission-line modelling.  
The OASIS survey that we present was designed preferentially for the study
of the gas dynamics, and the signal-to-noise ratio (S/N) 
of the acquired data was therefore calculated for optimal 
results in the gas emission lines. A simplified stellar-population modelling,
with the primary goal of correcting the Balmer lines of hydrogen for 
underlying stellar absorption, permitted us to investigate both the stellar 
kinematics and the age of stellar populations. 

Methods for the simultaneous fitting of stellar and 
emission gas spectra were recently developed by the SAURON consortium 
\citep[see e.g.,][]{Sarzi06,McDermid06} and others 
\citep[e.g.,][]{SanOrion}. We briefly tested this software on 
our data and found that the low S/N in our stellar lines prevented us from 
judging reliably which method was more suitable for our data.  
In addition, the two-stage spectral analysis, with the independent 
treatment of the stellar and gaseous components, is advantageous      
for detailed studies of complex emission-line profiles:  
the stellar component removed, different 
emission-line models with an increasing degree of precision  
can gradually be applied, as needed.

For each of the data-analysis steps, the large volumes of data 
require to a large extent automatic treatment, with a subsequent check of 
the results, and  
detailed modelling applied only to selected spectra. 
The current paper presents our first-order results, while detailed
analysis and interpretation of selected objects 
is postponed to future papers 
dedicated to individual objects in the sample.

\subsection{Stellar populations \label{stelpop}}

We modelled the stellar component of spectra using 
synthetic evolutionary models from the GALAXEV library of  
\citet{BC}. 
Understanding the stellar populations in Seyfert galaxies has 
always been a challenging problem \citep[see e.g.,][]{Cid05,Raimann03}, 
due to the presence of strong gas emission lines, which contaminate the 
stellar spectra.  A fairly standard approach has been to correct   
emission lines for stellar absorption by 
assuming that the nuclear stellar populations are identical 
to those in the outer parts of the host galaxy 
\citep[e.g.,][]{HoIII,BennertSey2}. However, evidence 
of young stars in the centres of Seyferts has been presented by e.g., 
\citet{Heckman97,Gonz98,Joguet01,Raimann03};      
and \citet{Cid04}, and therefore the use of purely old stellar populations
has been shown to be insufficient even for 
detailed gas-emission studies, because of the need to correct reliably 
for the effects of stellar absorption.  

Stellar-population modelling searches for a  
combination of template spectra from a theoretical or experimental library 
which reproduces best the observed galactic spectrum.  
As a typical inverse problem, the stellar spectrum reconstruction
has a number of free parameters (such as the ages and metallicities of 
the individual populations, and their mass fractions), which may not  
be possible to constrain to form a unique solution.  
Intrinsic degeneracies exist between age 
and metallicity, and they are further enhanced by 
measurement uncertainties and noise, which wash away the differences between 
spectrally similar components, 
as demonstrated by \citet{Cid04}. These authors showed for artificial data that
even an accurately fitting template does not ensure a  
correct physical interpretation. 
Nevertheless, a most robust determination of population fractions is 
possible if larger stellar-age bins are used  
(defined coarsely as young $<100$\,Myr, intermediate 100\,Myr --
1.4\,Gyr, and old $>1.4$\,Gyr in their study).

We adapted the stellar population study to our data
quality. We were limited by the S/N in stellar absorption lines,  
the relatively short
wavelength ranges with only a few prominent absorption lines,  
and contamination by gas emission lines.  
The fitting of the stellar was only possible in one of the two observed 
spectral
regions, i.e., the MR1 domain (see Sect.\,\ref{Obs}), which contains 
the \ion{Mg}{i}\,{\it b} complex at about $5170$\,\r{A}, and iron lines 
Fe\,5015\,\r{A}, Fe\,5270\,\r{A}. Other constraints to the models    
were provided by the absorption-line wings of {H$\beta$}. No such features
were detected in the MR2 domain, and therefore the stellar component 
of spectra 
in this domain was inferred by extrapolating the models fitted to MR1.

\begin{figure}[htb]
\begin{center}
\includegraphics[width=0.3\textwidth]{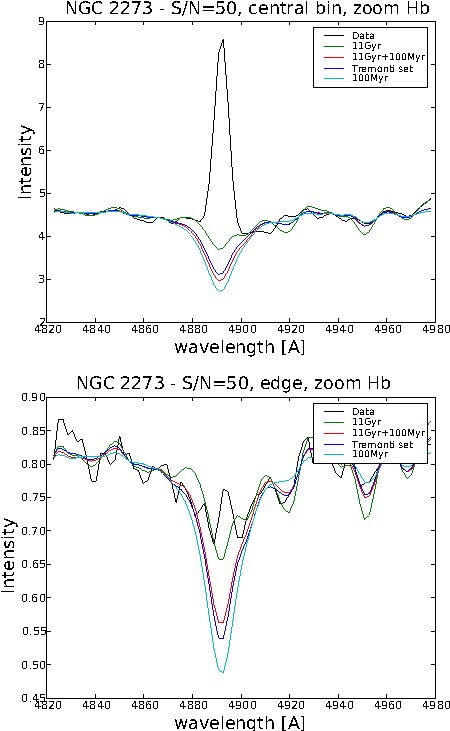}
\caption{\label{NGC2273SSPs} NGC 2273, comparison of SSP fits in the
central bin and a bin close to the FOV edge -- a young
stellar population (100\,Myr, cyan), an old SSP (11\,Gyr, green), an optimal
fit resulting from the combination of two SSPs: 11\,Gyr and 100\,Myr (red), 
and an optimal template based on the Tremonti set of SSPs (blue).
Spectra binned by Voronoi tessellation at $\mathrm{S/N} = 50$.
}
\end{center}
\end{figure}

From the GALAXEV models, we selected those that represent evolution
of Simple Stellar Populations (SSPs) with 
Salpeter initial mass function (IMF), 
by applying Padova evolutionary tracks \citep[e.g.,][]{Girardi00}. 
We based our work on the findings of \citet{Tremonti04}, and \citet{Cid04},
who tested different sets of SSP models for reconstructing 
spectra of active and non-active galaxies. 
They defined a spectral base of 30 SSPs covering ten ages and three 
metallicities, which played a role of independent functions, 
and any other SSP was found to be their linear combination.    
%
The tests that we performed with this ``Tremonti set'' on our data
revealed that the number of free parameters introduced by 
the theoretical combinations within the 30 SSPs was too large 
for our data quality, and   
could not be sufficiently constrained to produce a unique model, 
resulting in significant degeneracy with a difficult interpretation. 
We therefore searched for a simpler representation, 
and used models combining two SSPs of markedly different ages, 
as applied before e.g., by \citet{Canalizo07}. 
The even more basic option of single-population models 
proved to be inadequate -- the models failed to reproduce the observed 
spectral features 
correctly. The two SSP models were selected to be of solar metallicity, and 
of ages of 11\,Gyr and 100\,Myr, which were found to contribute 
the most even in the models to our data produced by the entire Tremonti set,  
and were able to describe many different types
of spectra across the whole sample. The final fits differ from those
obtained with the Tremonti set by up to 
$\sim\!\!20\%$ in the equivalent width of the {H$\beta$}
absorption line and by $\leq20\,$km\,s$^{-1}$ in the mean LOS 
stellar velocity (Fig.\,\ref{NGC2273SSPs}). 
We also studied other possible pairs of SSP models, but they were less 
successful
in reconstructing the {H$\beta$} absorption wings, the metallic lines, 
or predicting 
the {H$\beta$} absorption equivalent widths that 
differed significantly from the Tremonti set. Some SSP model 
combinations (such as those of ages  
11\,Gyr and 600\,Myr) proved useful for certain objects, but the best-fit
solution did not differ significantly from the final pair of SSPs 
(11\,Gyr and 100\,Myr), and did not provide  
optimal solutions for the remainder of the sample.

The best-fit model stellar template for each observed spectrum produced 
from a linear combination of selected SSPs was determined by using the 
penalised pixel-fitting code of \citet{Cappel}, and a careful 
masking of emission lines.  
For the purposes of inferring both stellar kinematics and stellar 
population ages reliably, we increased the signal-to-noise ratio 
by rebinning the FOV with equi-mass Voronoi tessellation \citep{Voronoi} 
into irregular compact cells of uniform  
$\mathrm{S/N} = 50$ across the FOV. 
For other applications employing the stellar model to correct 
the gas emission, we modelled all the individual observed spectra, without 
losing spatial resolution.

\subsection{Emission lines \label{emlinessect}}

The two observed spectral domains MR1 and MR2  
contain two Balmer lines ({H$\alpha$} and {H$\beta$}) and five 
forbidden-line doublets 
of metal ions that represent some of the most fundamental spectral diagnostics 
for deriving physical parameters of the ionised gas:  
[\ion{O}{iii}]$\,\lambda\lambda4959,5007,$
[\ion{N}{i}]$\,\lambda\lambda5198,5299,$
[\ion{O}{i}]$\,\lambda\lambda6300,6364,$
[\ion{N}{ii}]$\,\lambda\lambda6548,6583,$ and
[\ion{S}{ii}]$\,\lambda\lambda6717,6731.$
In addition to the intensity in individual lines, we derive their 
mean line-of-sight velocities and line-of-sight
velocity dispersions and plot their two-dimensional distributions.   
Not all of the targets were observed in both spectral 
domains (see Table\,\ref{list}). 
Where the data are available, we computed the electron density from the 
[\ion{S}{ii}] doublet and estimated the interstellar reddening from the 
ratio of Balmer lines \citep{Osterbrock89}.

We modelled all narrow emission lines using single Gaussian profiles, 
even though asymmetric lines were expected at many locations.  
Cases for which we obtained large FWHM values of Gaussian models 
led us to identification of spatial regions   
where multi-component fitting might be necessary.  
We used the {\tt fit/spec} line-fitting software \citep{Rousset92}, 
which has a core in C and interface in ESO MIDAS, 
and permits automated fitting over entire datacubes, 
with constraints applied    
to the line intensities, intensity ratios, velocities, and line widths.   
We restricted the number of free parameters by fixing the ratios of 
the line intensities within the doublets to comply with the rules of 
atomic physics: \\ 
$0.5<\mathrm{[\ion{N}{i}]}\,\lambda5198\,/\,
 \mathrm{[\ion{N}{i}]}\,\lambda5200<1.5;$ \\ 
$\mathrm{[\ion{O}{iii}]}\,\lambda5007\,/\,
\mathrm{[\ion{O}{iii}]}\,\lambda4959 = 2.88;$ \\   
$\mathrm{[\ion{O}{i}]}\,\lambda6300\,/\,
\mathrm{[\ion{O}{i}]}\,\lambda6364 = 3.0;$ \\ 
$\mathrm{[\ion{N}{ii}]}\,\lambda6583\,/\,
  \mathrm{[\ion{N}{ii}]}\,\lambda6548 = 2.96;$ \\    
$0.35<\mathrm{[\ion{S}{ii}]}\,\lambda6717\,/\,
   \mathrm{[\ion{S}{ii}]}\,\lambda6731<1.5\,.$

To constrain line parameters more effectively in the automated fitting 
process, especially for the weaker or blended lines, 
the emission lines were associated to form systems of assumed common 
kinematics. The mean LOS velocity and LOS velocity dispersion were thus 
determined for each of the systems, rather than the individual lines. 
The definition of our kinematic systems was physically and experimentally 
motivated. The kinematical properties of O$^{++}$ ions  
are known to differ from those of the low-ionisation species, 
due to the possibly different processes responsible for the ion formation. 
The first ionisation potential of elements such as oxygen, nitrogen, 
and sulphur is almost identical to that of hydrogen, and the ions 
can therefore be formed in the same conditions. And indeed, the 
observations 
show that kinematic properties of the neutral and low-ionisation species    
are usually similar to those of hydrogen.   
With this justification, we associated the weak and blended 
[\ion{N}{i}] doublet with {H$\beta$} in the same system, while 
[\ion{O}{iii}] was treated 
independently. All of the MR2 lines ({H$\alpha$}, [\ion{O}{i}], 
[\ion{N}{ii}], and [\ion{S}{ii}])
were considered in a common system (where possible) to assure 
a good quality fit to lines that are rather weak 
(mainly [\ion{N}{i}], [\ion{O}{i}], and [\ion{S}{ii}]) or blended 
([\ion{S}{ii}] doublet lines, [\ion{N}{ii}] with {H$\alpha$}). 

We corrected the measured FWHM for the instrumental dispersion, which 
was estimated by fitting telluric lines to be $(7\pm0.5)$\,\r{A}.  
A Gaussian approximation to the line profiles was used, hence with a quadratic 
subtraction of the measured total and instrumental dispersions.
The resulting dispersion contains an error of 0.6\,\r{A} due to wavelength
calibration. The differences encountered between the resultant 
velocity dispersion measured in {H$\alpha$} and in {H$\beta$} 
can be almost entirely 
attributed to their different data quality, far higher 
for {H$\alpha$}. Where available, {H$\alpha$} results should therefore 
be considered more reliable.

\subsection{Error computation}

Uncertainties in the measurements of the mean LOS velocity, which are a
function of the signal-to-noise ratio, were studied for OASIS data 
in detail by \citet{Ferr2110}, who fitted repeatedly artificial emission-line 
spectra with added noise in 500 Monte Carlo realisations for each 
S/N value \citep[see also][]{McDermid06}. 
Their results were applied in the way recommended by the authors 
in the present paper.    
We show the corresponding error bars in one-dimensional plots of
mean LOS velocities in Figs.\,\ref{Mrk34vel}~--~\ref{NGC5929maps}.
The effects of fitting single Gaussian functions, and of observed line-profile 
asymmetries were not included in this treatment.  
Equally, errors introduced by the subtraction of the stellar component,
especially in the hydrogen lines, were not included in the error bars.
An error of $16$\,km\,s$^{-1}$ due to calibration was added to the computed 
error in the mean LOS velocity. 

The random errors in line flux determination were estimated with the
assumption of Gaussian line profiles, and were computed as the product of 
noise in the vicinity of the lines, and the line FWHM. 
This procedure does not include possible systematic errors 
due to the stellar 
template determination, which are difficult to quantify, as 
discussed in Appendix B3 of \citet{Emsellem04}. 
We plot error bars in the scatter plots of
the derived quantities such as the line ratios and the one-dimensional 
cuts of electron densities, for all of which  
Gaussian error propagation was applied. 
For electron densities, the errors were computed assuming further 
that the [\ion{S}{ii}]
line shifts and FWHMs corresponded to those of {H$\alpha$}, 
i.e., we did not consider the effects of line blending.           

\section{Results for individual objects\label{results} }

\subsection{General comments}

We present colour-coded maps of the line emission intensity, 
the mean line-of-sight velocity, and the LOS velocity dispersion 
in the detected emission lines for each of the Seyfert\,2 galaxies  
in our sample. Thanks to our stellar-population modelling, the 
gas velocity maps can be compared with those of stars in the objects where 
the stellar data quality was sufficient for reliable modelling of the 
absorption-line profiles.  
The stellar modelling also yielded maps of the mass 
fractions of the young and old stellar populations. 

Further information about the objects can be inferred from the ratios of 
the emission-line intensities.  
We derive electron density from the [\ion{S}{ii}] line ratio, as described in
\citet{Osterbrock89},  
by assuming a temperature of $10\,000$\,K (our data do not contain a 
direct temperature diagnostic). 
We estimate the dust distribution from the Balmer line ratio. 
The {H$\alpha$/H$\beta$} should have a fixed intrinsic value depending 
on the physical conditions in the environment where the recombination 
lines are formed. 
We assumed the intrinsic value of 3.1, which is typical of 
NLRs \citep{Osterbrock89}. By assuming any excess of the measured ratios 
to be due to dust along the line of sight, we computed 
the interstellar reddening $E(B-V)$. 

The ratios of forbidden-line to recombination-line intensities 
are a measure of the 
ionisation and excitation state of the gas, of its density, temperature and
metallicity, as well as the nature of the ionizing
source. We plot maps for each of the standard diagnostic 
ratios [\ion{O}{iii}]$\lambda5007$/{H$\beta$} (excitation map),
[\ion{O}{i}]$\lambda6300$/{H$\alpha$}, 
[\ion{N}{ii}]$\lambda6583$/{H$\alpha$}, 
and ([\ion{S}{ii}]$\lambda6717$+[\ion{S}{ii}]$\lambda6731$)/{H$\alpha$}. 
More information is conveyed
by the spectral-diagnostic diagrams defined by \citet{BPT}; and 
\citet{Veill87}, but in our case, these diagrams are
plotted for spatially resolved data for each observed galaxy. The 
diagrams provide further possibilities for future detailed modelling of the 
ionisation structure, similar to that  
performed e.g., by \citet{Bennert1368}, or \citet{Allen99}
for long-slit spectroscopic data.    

In our diagnostic diagrams, we compare the different existing
classification schemes for emission objects commonly used in the literature,    
including the recent development \citep{Kew01,Kauf,StasiAGN,Kew06}, 
as well as the traditional schemes, some of which were based on a single 
pair of lines and were reviewed by \citet{HoIII}.  
However, we note that the models correspond to global 
spectra, and one must be careful when interpreting the 
spatially resolved diagrams \citep[see][]{StasiWS}, which require
detailed ionisation modelling for a proper understanding. 
Therefore, the plotted curves that correspond 
to the afore mentioned models play an informative role only.

All the plots presented in Figs.\,\ref{Mrk34maps}\,--\,\ref{NGC5929maps} 
show the results of single-Gaussian modelling of the emission lines. 
This simplification is clearly insufficient for a precise reconstruction of
many spectra, where
two or more components are present in each emission line, as specified 
below for the individual objects. 
The single-component fitting obviously limits the interpretation of the 
presented 
data, although the results provide a guide to 
the location of important structural and kinematic features in the objects.  
The applied modelling approach simulates the effects of lower resolution 
spectral data with unresolved 
multiple components. The resulting intensity in the lines corresponds to the
integrated intensity of all components or to the intensity in the
strongest component, depending on the line shape and the fit. 
Similarly, the fitted line-of-sight velocities trace either the motion of the 
strongest component at a given location or a ``weighted average'' of the 
components. The velocity-dispersion maps 
serve as efficient indicators of locations of the multi-component spectra. 
While it would be impossible to inspect all spectra individually 
due to the enormous amounts of data, 
the large-dispersion areas of the FWHM maps trace the regions of complex 
emission-line profiles, often in unexpected locations, not necessarily at the
galactic centres.

We display profiles of the mean LOS velocity and electron density
along the major axes of emission, and scatter plots of radial variations 
in the diagnostic line ratios as complementary information. 
Most of the data presented were clipped at $\mathrm{S/N}=5,$ with the
exception of the weak lines [\ion{N}{i}] and [\ion{O}{i}], for which 
only at a few spatial pixels in the centre 
the signal-to-noise ratio exceeded the value of two, and
therefore we applied clipping at $\mathrm{S/N}=1$ 
to plot the two-dimensional maps.
The surface brightness is in the units of 
$10^{-19}$\,J\,s$^{-1}$\,arcsec$^{-2}$\,m$^{-2}$, 
the mean LOS velocities, and LOS velocity dispersions in km\,s$^{-1}$.
The velocities were corrected for heliocentric motion using the
{\tt rvcor} routine of IRAF, and plotted relative to the systemic
velocity of each galaxy, which was taken from NED (see Table\,\ref{physpar}).

\bigskip

We compared the measured fluxes of emission lines with 
those determined by 
\citet{Falcke98}. The authors provide the fluxes in [\ion{O}{iii}] and 
{H$\alpha$}+[\ion{N}{ii}]
obtained from HST observations of {Mrk 34}, {Mrk 348}, 
and {NGC 4388},
using circular apertures of two different sizes for each object.
We simulated the apertures by summing the spectra over
circular regions centred on $[0\arcsec,0\arcsec]$ and of the same radii
as in the cited paper.
The emission-line fluxes were measured for spectra that had a 
stellar component modelled by a combination of two SSPs (11\,Gyr and
100\,Myr) from \citet{BC}, as described in Sect.\,\ref{stelpop}, and with
no emission-line model applied (to avoid introducing further sources of
error).
Results of the comparison are summarised in Table\,\ref{tabflux}.

Agreement between the two data sets is closer for larger apertures, 
as expected.
The flux that we measure in regions of radii smaller than $1\arcsec$ is
affected too much 
by the spatial PSF (dominated by $\sim\!\!1\arcsec$ seeing), which is 
an order of magnitude higher than that in the HST observations and
redistributes much of the observed flux outside the selected area.
Most of the fluxes that we obtained for apertures of sizes
$\geq1\arcsec$ are $10\%-25\%$ 
lower than previously published values.
Possible reasons for the discrepancy include flux-calibration errors, 
errors acquired during data reduction or due to an imprecise
determination of the stellar-continuum level. 
However, it is impossible to draw conclusions about systematic differences
based on only three cases. 

\begin{table}[hbt]
\caption{\label{tabflux}
\small Comparison of our emission fluxes with those of \citet{Falcke98}. 
OASIS values
correspond to fluxes corrected for absorption using a combination of SSPs as
described in Sect.\,3.1.  Units are  $10^{-19}\,\mathrm{W\,m}^{-2}$.
}
\begin{center}
\tiny
\begin{tabular}{|ll||l|l||l|l|}
\hline
Galaxy  & Aperture & \multicolumn{2}{c||}{OASIS data} &
\multicolumn{2}{c|}{\citet{Falcke98}}\\
\cline{3-6}
& radius   & [\ion{O}{iii}]& {H$\alpha$}+[\ion{N}{ii}] & 
[\ion{O}{iii}]& {H$\alpha$}+[\ion{N}{ii}]\\
\hline
\hline
Mrk 34  & 2\arcsec      & 4\,800 & 3\,400  & 6\,100 & 3\,600 \\
& 0.5\arcsec    & 500   &  300    & 1\,300  & 700 \\
\hline
Mrk 348 & 1.5\arcsec    & 3\,000   & 2\,500  & 4\,100  & 2\,000 \\
& 0.4\arcsec    & 1\,500   & 1\,000  & 3\,200  & 1\,500 \\
\hline
NGC 4388& 6\arcsec      & 9\,800   & ---  & 11\,700    & --- \\
& 1\arcsec      & 3\,200   & ---  & 3\,900     & --- \\
\hline
\end{tabular}
\end{center}
\end{table}

\bigskip

The following subsections are devoted to the description of the individual
objects, the observed features, set into the larger multi-wavelength context
found in the literature. We note that we will not be restating the wavelengths 
of the emission lines that were listed in Sect.\,\ref{emlinessect}, 
except the  cases where confusion would be possible.   
 


\subsection{Individual objects}

\subsubsection*{{Mrk 34}}

{Mrk 34} is known to be a luminous FIR source with a powerful 
water maser \citep{HenkelMrk34}.   
Optical line emission was found  to be spatially correlated with 
radio continuum emission, $\mathrm{PA} \sim 150\degr - 165\degr$
\citep{Nagar99Orient,Falcke98,Ulve84V}. 

\bigskip

{Mrk 34} is the most distant object in our OASIS sample 
and in this case
our FOV corresponds to the central $\sim\!\!5$\,kpc. 
Because of the high redshift of the galaxy, the [\ion{S}{ii}]
doublet is located 
outside the observed spectral range, and we are therefore unable to 
infer all of the spectral-diagnostic ratios, and the electron density.  
The observed emission in all of the observed 
lines has a similar morphology, extended along the direction of 
$\mathrm{PA} \sim 140\degr-150\degr$ (Fig.\,\ref{Mrk34maps}). 
The excitation map 
[\ion{O}{iii}]/{H$\beta$} has the same major axis as the
[\ion{O}{iii}] emission, and more of a cone-like shape. 
The other diagnostic line ratios, however, have low values in the direction
of the cones, while the maximum values are distributed along the minor axis 
(Fig.\,\ref{Mrk34maps}). 
Reddening computed from the {H$\alpha$}/{H$\beta$} ratio possesses a maximum
in the southern part of the FOV, $\sim\!\!2$\,kpc from the central maximum.
The spatially resolved diagnostic diagrams indicate that the emission 
corresponds to a strong Seyfert regime.

The map of the mean LOS velocities in {H$\alpha$} 
(Fig.\,\ref{Mrk34vel}) provides an  
example of twisted S-shape isocontours. They are not evident in 
[\ion{O}{iii}] 
and {H$\beta$}, which may partially be caused by the smaller FOV in MR1. 
No galactic bar or other signature of non-axisymmetric potential 
has been reported so far, and the S-shape might be the first piece of
kinematical evidence of departures from axial symmetry. 
The low quality of the stellar data did not allow comparison of the stellar 
and gas velocities.  
The object is characterised by multi-component emission-line profiles 
in a region of elongated shape perpendicular to the emission major axis, 
as visible in the velocity dispersion map.

\subsubsection*{{Mrk 622}}

The radio emission in this galaxy at the 20\,cm wavelength is extended 
along $\mathrm{PA} =0\degr$ \citep{Nagar99Radio}. 
No prominent axis in either the optical emission or the excitation map was 
detected on the scale of $5\arcsec$ by \citet{MulObs},  
while \citet{Schm03Obs} reported elongation in [\ion{O}{iii}] emission at 
$\mathrm{PA} =55\degr$ in the inner $1\arcsec,$  
perpendicular to the host-galaxy major axis.
\citet{Shuder81} discovered multi-component [\ion{O}{iii}] profiles of 
total $\mathrm{FWHM} = (1050\pm50)$\,km\,s$^{-1}$, in contrast to the
nearly-Gaussian shapes of the {H$\alpha$}, [\ion{N}{ii}], 
and [\ion{O}{ii}] lines of 
$\mathrm{FWHM} = 350\pm 75$\,km\,s$^{-1}$. They concluded 
that two emission-line regions of different ionisation are present, with
estimated temperatures of $\sim\!\!11\,000$\,K 
in [\ion{O}{iii}] and $\sim\!6000$\,K in H$\alpha$, [\ion{N}{ii}], 
and [\ion{O}{ii}].

\bigskip   

The galaxy is among the most distant in our sample, which has an 
effect on the spatial resolution that we attain.
The maps of line emission intensities do not show any particular
directionality, except for the {H$\alpha$} and [\ion{N}{ii}] emission, 
which has hints of 
extending both to the east and the south (Fig.\,\ref{Mrk622maps}),  
in agreement with 
the optical and radio observations of \citet{Schm03Obs}; and 
\citet{Nagar99Radio}.
The [\ion{O}{iii}] lines are highly asymmetric and their multiple 
components are detected  throughout the field of
view (Fig.\,\ref{Mrk622spec}), with a separation between the components
as large as 600\,km\,s$^{-1}$, not present in the other lines, 
confirming the results of \citet{Shuder81}. 
Single-Gaussian modelling of [\ion{O}{iii}] thus results in large-FWHM fits,
as seen in Fig.\,\ref{Mrk622vel}.  

The one-component fits have a strong impact on the 
interpretation of the velocity field of this object. 
The relative intensity of the two
line components in [\ion{O}{iii}] varies across the FOV, with the
blue component becoming stronger in the east and the red component in
the west. The fitted one-component Gaussian model of the line profile is
always biased towards the stronger component. 
As a result, our velocity map in [\ion{O}{iii}] 
is not a good approximation of the true motion and a detailed 
multi-component modelling of line profiles is necessary.  
We assume that this is the major reason for the 
dissimilarities between the hydrogen and [\ion{O}{iii}] kinematics 
presented in Fig.\,\ref{Mrk622vel}.    
While the apparent kinematic axis of [\ion{O}{iii}] lies along 
$\mathrm{PA} \sim 90\degr$, that of {H$\alpha$} (together with the 
low-ionisation lines) 
is perpendicular to it, at $\mathrm{PA} \sim 180\degr,$ measured 
in the outer parts of the {H$\alpha$} map, the central regions being 
characterised by S-shape isovelocity 
contours. The stellar velocity field, as far as can be deduced from 
our data, has a kinematic axis at $\mathrm{PA} \sim 120\degr,$ and it is 
impossible to decide whether the twisted contours are present due to the low
data quality.  

The {H$\beta$} data are of low quality and their  
derived velocities do not agree with those measured from {H$\alpha$}. 
We note that velocity colour scales in Fig.\,\ref{Mrk622vel} were not 
homogenised to the same limits, due to the problematic fitting results.     
Two-component line profile modelling is necessary for [\ion{O}{iii}]
lines, as the ambiguous interpretation of present results demonstrates.

The electron density, the interstellar reddening, and the gas excitation 
all reach their maximum values at the galactic centre.  
In contrast, the  
diagnostic ratios of line intensities, except for 
[\ion{O}{iii}]/{H$\beta$}, reach their minima in the central region. 
Interestingly, the spatially resolved diagnostic diagrams 
indicate low excitation across the entire FOV (Fig.\,\ref{Mrk622maps}). 
The observed spectra possess strong {H$\beta$} absorption wings, which 
affect both the selection of the stellar population model and the 
predicted absorption in {H$\beta$}, which may be overestimated.

\subsubsection*{{Mrk 1066}}

While being classified as a marginal Seyfert by \citet{Oster83}, 
a high-excitation component of the spectral lines was found by 
\citet{Bower95}.
Radio emission was observed at $\mathrm{PA}  \sim 135\degr$ by 
\citet{Haniff}; and \citet{Ulve89}, and the narrow-band images 
indicated an alignment of the optical [\ion{O}{iii}]  
emission with the radio jet \citep{Bower95,MulObs}. 
The direction of both types of emission 
then agrees approximately with the orientation of the
galactic bar of $\mathrm{PA} \sim 143\degr$ \citep{Mazza93,Bower95}. 
According to \citet{Bower95}, the velocities in {H$\alpha$} + [\ion{N}{ii}] 
are consistent
with rotation, probably in the galactic disc, while [\ion{O}{iii}] 
shows little or no rotation, which they interpret as an outflow. 
No signs of hidden BLR were inferred from spectropolarimetry 
\citep{Miller90}. No central dust lane in 
high-resolution HST images, nor other 
evidence of a torus (such as excess reddening at the centre) was 
found by \citet{Bower95}. \citet{Zhang} attributed the non-detection of a
hidden BLR to orientation effects and/or the low mass of the 
central black hole.

\bigskip

The emission that we observe in all the lines is slightly extended along
$\mathrm{PA} \sim130\degr-140\degr$, i.e., parallel to the galactic bar.
A prominent emission feature located in the east is seen in the hydrogen
lines and partially in [\ion{N}{ii}] and 
[\ion{S}{ii}], while absent in [\ion{O}{iii}], 
and is associated with high interstellar reddening, low excitation, and a 
small velocity dispersion (Figs.\,\ref{Mrk1066maps} and \ref{Mrk1066vel}). 
We conclude that it is probably a region that has not been ionised by the AGN. 
Excitation is highest in the central parts and extends asymmetrically 
to the north-west side along the major axis of [\ion{O}{iii}] 
emission, as well 
as along the minor axis to both sides from the centre. Interestingly, 
the interstellar reddening extends asymmetrically along the 
major axis in the opposite direction, to the south-east 
(Fig.\,\ref{Mrk1066maps}). The maps of diagnostic ratios 
[\ion{O}{i}]/{H$\alpha$} and 
[\ion{N}{ii}]/{H$\alpha$} exhibit surprising 
distributions of high values within an 
extended region along the minor axis (Fig.\,\ref{Mrk1066maps}). The 
diagnostic diagrams are consistent with the interpretation of the
object as a marginal Seyfert (Fig.\,\ref{Mrk1066maps}). 
With a few exceptions, all of
the spectra are characterised by an [\ion{O}{iii}]/{H$\beta$} 
intensity ratio that is 
below the empirical Seyfert limit, including the central parts.
However, a proper disentangling of the asymmetric lines  
may confirm the high-excitation component (\citep{Bower95}). 
The electron density reaches peak in the central region, the maximum 
being off-centred by $\sim\!\! 200$\,pc (Figs.\,\ref{Mrk1066vel} and 
\ref{Mrk1066spec}).         

The {H$\alpha$} velocity map (Fig.\,\ref{Mrk1066vel}) displays a clear 
S-shaped pattern. 
Velocities in [\ion{O}{iii}] exhibit important differences from the
hydrogen data, seen mainly in the one-dimensional cuts. 
We found no multi-component line profiles for this object
(Fig.\,\ref{Mrk1066spec}), even though we observe a 
large-FWHM region extending along the minor axis, 
which is usually associated with line splits in other objects of the sample.
On the other hand, most lines have blue wings across the entire FOV.

\subsubsection*{{NGC 262 }(Mrk 348)}

The host-galaxy is orientated almost face-on and has a close companion, 
NGC~266. The nuclear region has been a popular target of observations 
across the entire electromagnetic spectrum, because of 
the difficulties to interpret orientations of individual components     
observed in the different spectral regions 
within a simple model \citep{Anton02}.      
\citet{CapettiRadio} reported a prominent central dust lane of  
scale height 50\,pc, perpendicular to the optical [\ion{O}{iii}] emission 
in an HST image. Evidence of a torus or 
related structure was found in IR images by \citet{Simpson96}.   
A polarised broad {H$\alpha$} component \citep{Miller90} and a hard X-ray 
emission \citep{Awaki} inferred a hidden Seyfert\,1 nucleus.  
The radio emission is in the form of a triple radio source of spatial extent 
$0.2\arcsec$ along $\mathrm{PA} =168\degr$  \citep{Neff}, variable 
on the scale of months. 
Megamaser emission detected in the central few
milliarcseconds \citep{Falcke00,Xan,Peck03} was interpreted as an 
interaction between the radio jet (situated close to the plane of the sky)
and a molecular cloud. 
Optical [\ion{O}{iii}] emission is well aligned with the radio 
on sub-arcsecond scales \citep{Falcke98,Schm03Obs}, 
while on the arcsecond scales 
it is extended along $\mathrm{PA} \sim 185\degr$ \citep{Schm03Obs}, 
with a secondary blob of emission $\sim\!0.9\arcsec$ south of the nucleus.    
A hint of ionisation cones was found by 
\citet{Simpson96}; and \citet{MulObs} in the ionisation map, 
but the evidence is marginal according
to \citet{Anton02}.   
NIR emission map obtained with an IFU by \citet{Sosa} does not have any
prominent direction on the scale of $3\arcsec$, although contours in the inner
parts seem to have a slight directionality along $\mathrm{PA} 
\sim 10\degr$ and $\mathrm{PA} \sim90\degr.$  
A ring of \ion{H}{ii} emission on the scale of $\sim\!\!1$\,kpc was reported by
\citet{Anton02}; and \citet{MulObs}.
The galaxy possesses a giant \ion{H}{i} envelope on the scales
of hundreds of kpc \citep{Heckman}.

\bigskip

The face-on 
orientation of the galaxy is confirmed in our data by the low stellar
LOS velocities (Fig.\,\ref{Mrk348vel}), the minimum and maximum values 
that we obtained from our fits are separated by $70$\,km\,s$^{-1}$ only. 
We estimate the kinematic stellar axis to be at $\mathrm{PA} 
\sim 70\degr.$
This Seyfert galaxy has one of the lowest contributions 
from young stars of the entire sample. Mass fractions of young 
stars above several percent have been identified in the central 
$\sim\!\!250$\,pc only. 

The emission maps of most of the emission lines are  
slightly extended along $\mathrm{PA} \sim10\degr$ (Fig.\,\ref{Mrk348maps}),
which is consistent with \citet{Sosa}, and differs by approximately 
$20\degr$ from the radio jet reported in the literature. 
We partially resolve the secondary emission maximum known from the literature, 
$\sim\!\!1\arcsec$ south-west of the centre. 
The gas excitation is high, and all the regions are high in the Seyfert
regime according to the spatially resolved diagnostic diagram 
(Fig.\,\ref{Mrk348maps}). The regions of highest excitation are situated along 
the major axis of emission, within a $\mathrm{PA} $ range of 
$\sim5\degr - 30\degr\,.$
The [\ion{O}{iii}]/{H$\beta$} ratio decreases steeply with radial distance, 
as seen in the corresponding graph of Fig.\,\ref{Mrk348vel}.  
We note that the ratios of low-ionisation lines to {H$\alpha$} show 
prominent minima south-west of the galactic centre, in the region of the 
receding velocity maximum (see paragraph below).    
In addition, [\ion{S}{ii}]/{H$\alpha$} has a minimum at the centre, 
surrounded by a patchy ring of high [\ion{S}{ii}]/{H$\alpha$} on scales of 
$\sim\!\!600$\,pc, which corresponds to a ring of 
low electron densities derived from the [\ion{S}{ii}] ratio 
(Figs. \ref{Mrk348maps} and \ref{Mrk348vel}).

The gas velocities (Fig.\,\ref{Mrk348vel}) are the most 
surprising detection. 
We observe two spots of velocity maxima of opposite sign, aligned at 
$\mathrm{PA} \sim25\degr,$ i.e., approximately parallel to the major 
axis of optical emission.  
The two spots of $>100$\,km\,s$^{-1}$, 
present in [\ion{O}{iii}] as well as in low-ionisation lines, 
about $200$\,pc in size,
are placed symmetrically $\sim\!\! 300$\,pc from the emission maximum.  
We assume that the observation corresponds to a rotating ring, inclined 
with respect to the galactic disc.    
However, there is a clear asymmetry in the radial profile of the
two velocity peaks (Fig.\,\ref{Mrk348vel}).

The maxima of the LOS velocity dispersion in {H$\alpha$} are located 
at the positions of the velocity peaks, which is not the case for
 [\ion{O}{iii}].
The low-ionisation emission lines also have 
significantly asymmetric profiles across the FOV, unlike
[\ion{O}{iii}] where mostly blue wings are present (Fig.\,\ref{Mrk348spec}). 
The low-ionisation lines are characterised by 
prominent blueshifted and redshifted components 
shifted by $300-400$\,km\,s$^{-1}$ from 
the main component. The red asymmetries were found mainly in the 
southern part of the FOV. The vicinity of the south-western spot of 
maximal velocities is characterised by rapid spatial variations 
in the line shapes and 
switching between red and blue additional components. 
A careful disentangling of the line profiles is necessary
before further conclusions can be made.    

We detect another systematic motion in [\ion{O}{iii}], 
at the edge of the FOV, 
of radii $\sim\!\!1$\,kpc, where we observe velocities with a polarity 
that is reversed
with respect to the nuclear ring. We verified that it is 
not the effect of noise and suggest that it might correspond to the ring
identified by \citet{Anton02}; and \citet{MulObs}.

\subsubsection*{{NGC 449} (Mrk 1)} 

The galaxy is well known to possess a prominent water maser in its
nucleus \citep{Braatz94}. No evidence of a hidden Seyfert\,1 
nucleus was found, in either
polarised light \citep{Kay94} or in the infrared \citep{VeillIR}.  
\citet{MulObs} reported [\ion{O}{iii}] emission extending 
to $\sim\!\!3$\,kpc, 
co-aligned with both the galaxy major axis, $\mathrm{PA} =83\degr$ 
and 20\,cm radio emission on the scale of $\sim\!\!30$\,pc 
\citep{Kukula99}.    
Stellar-population synthesis modelling of 
long-slit data of 3\,\r{A} resolution
performed by \citet{Raimann03} revealed mostly  
intermediate-age stellar populations ($\sim\!\!1$\,Gyr) in the 
central kpc and an
increasingly dominant old population ($\sim\!\!10$\,Gyr) 
outside this radius.

\bigskip

Our OASIS data for this object were obtained as a mosaic of 
three partially overlapping fields. 
We confirm strongly collimated emission in [\ion{O}{iii}], 
\ion{H}{ii}, and [\ion{N}{ii}] 
extending to radii of at least $\sim\!\!3$\,kpc (Fig.\,\ref{Mrk1maps}). 
In the other observed 
lines, the S/N is insufficient to obtain reliable data to these distances.   
We found a misalignment between the emission in 
[\ion{O}{iii}] ($\mathrm{PA} \sim65\degr$) 
and the low-ionisation species ($\mathrm{PA} \sim85\degr$). 
The velocity map in {H$\alpha$} suggests circular rotation.   
As in most of the targets of the OASIS sample, we found a
region of large FWHM, stretching along the minor axis (Fig.\,\ref{Mrk1vel}). 
The region is characterised by multi-component line profiles, usually in the 
form of red or blue wings in [\ion{O}{iii}] and split line profiles in 
low-ionisation species (Fig.\,\ref{Mrk1spec}).  
Outside the central $\sim\!\!600$\,pc, the emission lines are narrow.    

Our stellar data were not of a high quality for this object. The maximum 
contribution of young stars was fitted in the nuclear region. The 
map of stellar velocities could not be interpreted reliably.

\subsubsection*{{NGC 2273} (Mrk 620)}

{NGC 2273} is a barred galaxy \citep{vanDriel91},  
with a star-forming nuclear ring \citep{FerrHST}.
Deviations from axial symmetry were confirmed 
by GMOS IFU observation of the stellar kinematics in 
the central regions \citep{Barbosa}.
Radio emission extends along $\mathrm{PA} \sim20\degr$ at the 
20\,cm wavelength, 
and two components along $\mathrm{PA} \sim90\degr$ are evident at 
the 6\,cm wavelength in the central $1\arcsec$ \citep{Ulve84,Ho01}.
HST observations resolved a jet-like [\ion{O}{iii}] structure 
extending $2\arcsec$ east 
of the nucleus, at $\mathrm{PA} \sim90\degr,$ aligned with the double 
radio source \citep{FerrHST}.  \citet{MulObs} reported 
extended [\ion{O}{iii}] emission reaching to $\sim\!\! 6.5\arcsec$ at 
$\mathrm{PA} =134\degr\pm10\degr.$
\citet{Moiseev04} identified important differences between the velocities 
in [\ion{O}{iii}] and the Balmer lines.

\bigskip

Our data for this object are limited to a single spectral domain (MR1).    
While the [\ion{O}{iii}] emission does not exhibit any prominent 
axis (only a slight 
elongation is visible along $\mathrm{PA} \sim95\degr$), 
{H$\beta$} emission seems to
follow the direction of $\mathrm{PA} \sim20\degr.$ These differences 
in behaviour
are emphasised in the [\ion{O}{iii}]/{H$\beta$} map, where 
extended cone-like
features appear along $\mathrm{PA} \sim100\degr$ (Fig.\,\ref{NGC2273maps}).   
The velocity field in [\ion{O}{iii}] lines shows S-shaped isocontours 
(Fig.\,\ref{NGC2273maps}), which might
indicate deviations of the gravitational potential from axisymmetry.   
The data quality in {H$\beta$} is insufficient to confirm 
or exclude a similar velocity pattern in hydrogen. 
However, one-dimensional cuts of the velocity fields show important
systematic differences between LOS velocities in [\ion{O}{iii}] 
and {H$\beta$}.  
The stellar velocity field is well measured for this galaxy and shows 
rotation in a direction similar to gas, with the kinematic axis 
at the outer parts of the FOV
approximately parallel to the {H$\beta$} photometric major axis. 
No asymmetric or multi-component emission-line profiles 
have been found, apart from weak, blue wings (Fig.\,\ref{NGC2273maps}).

\subsubsection*{{NGC 2992}}

The host galaxy disc is highly inclined at 
$i\sim70\degr,$ which makes it suitable for extraplanar emission studies.  
The galaxy has a companion NGC 2993, which is located $\sim\!\!3\arcmin$ 
south-east,
connected with a tidal bridge, and whose tidal forces might have an important
effect on the kinematics.
Radio observations at 20\,cm found a $25\arcsec$ structure along 
the major axis of the
galaxy, and a $90\arcsec$ one-sided extension at 
$\mathrm{PA} \sim100\degr-130\degr$, 
close to the galaxy minor axis \citep{Ward}. The 6\,cm radio emission 
is concentrated in the central parts, forming a ``figure-eight'' structure
at $\mathrm{PA} =160\degr$ \citep{Ulve84,Wehrle88}. 
Soft X-rays are detected up to scales of $35\arcsec-45\arcsec$ 
along the galaxy's minor
axis, and are co-spatial with the 20\,cm radio emission.
The [\ion{O}{iii}] emission forms two sharp-cut cones with a 
projected opening angle of $120\degr$ 
\citep{Marquez98}, their axes aligned with the galactic minor axis 
($\mathrm{PA} =120\degr$) and approximately with the radio structure. 
A disturbed dust lane was identified at $\mathrm{PA} =15\degr$ by 
\citet{Marquez98} 
and at $\mathrm{PA} =30\degr$ by \citet{Veilleux2992}.
Multi-component optical emission lines were detected,   
and the velocity field was suggested 
to be a superposition of rotation and an outflow
\citep{Marquez98,Allen99,Veilleux2992}.

\bigskip

The FOV in OASIS observations enables  
only one of the emission cones to be imaged, located south-east 
of the dust lane. 
The emission structure that we observe (Fig.\,\ref{NGC2992maps}) is elongated 
at $\mathrm{PA} \sim30\degr$ in all of the
available emission lines, parallel to the dust lane observed by 
\citet{Veilleux2992}. 
The emission bends in the southern part towards $\mathrm{PA} \sim210\degr$, 
especially in
[\ion{O}{iii}], and to $\mathrm{PA} \sim45\degr$ in the north of the FOV. 
[\ion{O}{iii}] has a significant emission extension in the eastern 
part of the FOV, associated with high receding velocities,  
while the low-ionisation lines exhibit a blob of emission to the south-west,
associated with high extinction. 

The maximum of the emission in all lines coincides with the maxima in  
the electron density, the interstellar reddening, and the 
velocity dispersion (with the 
exception of [\ion{O}{iii}] where the FWHM maximum is shifted 
by $\sim\!\!1\arcsec$ to the east, see Fig.\,\ref{NGC2992vel}).    
At the same time, the maximum emission region is characterised by 
local minima of diagnostic line ratios (Fig.\,\ref{NGC2992maps}).

The LOS velocity fields (Fig.\,\ref{NGC2992vel}) obtained from 
different emission lines are similar,
characterised by receding motions in the northern part and approaching in the
south and east of the FOV.
Comparison with the stellar velocities (with their kinematic axis 
in the NE-SW direction) 
suggests that the gas has a rotational component 
aligned approximately with stars. 
Despite the apparent kinematic similarities, the MR2 spectra
had to be considered to correspond
to two kinematic systems, by fitting the [\ion{S}{ii}] and [\ion{O}{i}] 
kinematics
independently of {H$\alpha$} and [\ion{N}{ii}], while the unified fit 
was impossible. The grouping of the emission  
lines into the two systems was a compromise between the well-determined
positions of lines and the number of parameters needed to place enough 
constraints on the weak lines and the blended lines. 
We illustrate the blending of the {H$\alpha$} and [\ion{N}{ii}] lines 
in the spectra of Fig.\,\ref{NGC2992spec}. 
Outside the central $\sim\!\!2\arcsec$ (350\,pc), all 
emission lines are however narrow and non-blended. 
The differences in the motion of the individual ionic species
are depicted in the one-dimensional cut along the major axis
(Fig.\,\ref{NGC2992vel}): the
velocity curve in [\ion{O}{i}] and [\ion{S}{ii}] is shallower 
(by $\sim\!\!100$\,km\,s$^{-1}$) 
than in the other lines, whereas the most prominent 
velocities are observed in [\ion{O}{iii}], especially in the northern part. 
The velocity curves that we obtained by one-dimensional 
cuts along different axes 
are consistent with long-slit spectroscopy by \citet{Marquez98}. 

We detect double-component profiles of emission lines, particularly 
for [\ion{O}{iii}] and {H$\beta$}, while the other lines are too 
affected by either   
blending or noise. We show the asymmetric MR1 profiles in 
Fig.\,\ref{NGC2992spec},
in a combined plot of spectra from three locations. 
Broad wings of Balmer emission lines were detected in the central 
$\sim\! 2\arcsec$ 
($\sim\! 350$\,pc). However, their relation to BLR emission
\citep[already suggested e.g., by][]{Ward,Boisson86} can only be 
confirmed after a careful disentangling of all the line components.

Our stellar population modelling revealed spatial variations in the 
contribution of young stars, which is highest in the 
strongly emitting region close to the dust lane, 
while mostly old populations are present in the rest of the
FOV (eastern part).

\subsubsection*{{NGC 3081}}

The host galaxy is an early-type spiral with a weak large-scale bar and 
a nuclear
bar \citep{Buta90}, and four well-defined resonance rings:
nuclear, inner and two outer ones \citep{Buta98}.
The brightest emission-line region inside $2\arcsec$ has a linear 
morphology, well 
aligned with the radio emission \citep{Nagar99Radio}. 
Slitless spectroscopy by \citet{Ruiz} revealed two major kinematic components
in the [\ion{O}{iii}] line at $\sim\!\! -250$\,km\,s$^{-1}$ 
and $~ +50$\,km\,s$^{-1}$.

\bigskip

We observed the object in one spectral domain (MR1) only.  
The photometric axis in both [\ion{O}{iii}] and {H$\beta$} stretches along 
the $\mathrm{PA} \sim155\degr-165\degr,$ even though it is less pronounced 
in [\ion{O}{iii}] (Fig.\,\ref{NGC3081maps}). 
Unlike in [\ion{O}{iii}], the emission in {H$\beta$} is elongated 
asymmetrically to the south-east. 

The mean LOS velocity fields of [\ion{O}{iii}] and {H$\beta$} 
(Fig.\,\ref{NGC3081maps}) 
show similar structures, forming an S-shape in [\ion{O}{iii}].  
However, important differences between [\ion{O}{iii}] and {H$\beta$} 
velocities 
are encountered in one-dimensional cuts of the velocity maps.  
The S-shape is not reproduced in the stellar velocities,  
derived from high-quality absorption-line data 
in this object, which have a rotational pattern with  
a kinematic axis $\mathrm{PA} \sim90\degr$. The anomalous 
values of stellar velocities detected in several bins of the map in 
Fig.\,\ref{NGC3081maps} do not correspond to true velocities, 
and are results of poor fits in these particular positions. 
Interestingly, the northern region of receding [\ion{O}{iii}] velocities is
characterised by {\em minimum} values of the line width.
We found no prominent asymmetries in emission line profiles  
(Fig.\,\ref{NGC3081maps}). 

Our stellar population modelling revealed low contributions from 
young stars, compared to the other Seyfert nuclei in the studied sample
(Fig.\,\ref{NGC3081maps}). Nevertheless, the SSP ages form a clear spatial 
structure with a boundary at $\sim\!\!800$\,pc separating older populations 
inside this radius and younger outside.

\subsubsection*{{NGC 4388}}

The galaxy belongs to the Virgo cluster \citep{Phillips82,Yasuda}. 
The galactic disc of $\mathrm{PA} \sim90\degr$ is inclined 
at $i\sim\-78\degr$ \citep{VeilleuxBar4388}, with the 
north rim on the near side. The nucleus is a strong X-ray source 
\citep[e.g.,][]{Hanson90,Takano91,Lebrun92,Iwasawa97}. 
Radio-emission morphology suggests a collimated AGN-driven outflow reaching 
$\sim\!200$\,pc south of the nucleus \citep[e.g.,][]{Stone,Falcke98}. 
Weak broad {H$\alpha$} emission implies that a hidden type-1 nucleus is present
\citep{Filippenko}. 
Extended optical emission has two components of different 
excitation, one associated with the galactic
disc and another reaching $50\arcsec$ (4\,kpc) above the galactic 
plane in the form of two opposite cones
\citep[e.g.,][]{Pogge4388,Corbin,Falcke98,Veilleux4388}.
While \citet{Rubin4388} argued that anomalous kinematics in the central parts
of the galactic disc were indicative of presence of a discrete, rapidly 
rotating circumnuclear disc, 
\citet{VeilleuxBar4388,Veilleux4388} found that the gas kinematics were 
consistent with elliptical streaming in
the bar potential, plus a bipolar outflow extending out of the galactic
plane (and excluded other models for the extragalactic component).  
They argued that the bar-induced non-circular motions may account for the
emission-line splitting of the maximum amplitude of 
$\sim\!\!150$\,km\,s$^{-1}$ detected
symmetrically out to radii of $\sim\!\!10\arcsec$ ($\sim\!\!1$\,kpc) 
along the disc, i.e., close to the end of the bar.  

\bigskip

In the case of this nearby galaxy, our FOV is restrained and covers the
central $\sim\!\!400$\,pc only, which is insufficient for interpreting the 
large-scale structures known from e.g., \citet{Veilleux4388}, but we sample 
the central region with the spatial scale of $\sim\!80$\,pc, obtaining thus 
a highly detailed picture. 
The maximum of emission within our FOV is located in the south-west  
(Fig.\,\ref{NGC4388maps}), and the emission intensity 
isocontours have strongly irregular 
forms, pointing to the north-east $\mathrm{PA} \sim30\degr-45\degr$, i.e.,
a direction consistent 
with the extragalactic component reported by \citet{Veilleux4388}. 
The {H$\beta$} morphology differs slightly from [\ion{O}{iii}] 
in the norther part of the FOV, 
where it bends by $\sim90\degr$ at $\sim\!\!200$\,pc. 
The map of [\ion{O}{iii}]/{H$\beta$} ratio 
reveals high ionisation in the region of maximum [\ion{O}{iii}] 
emission extending
along $\mathrm{PA} \sim45\degr$ (Fig.\,\ref{NGC4388maps}). 

Due to good-quality stellar data in this object, 
our map of mean LOS stellar velocities exhibits a clear 
pattern, which is similar to 
that observed in {H$\alpha$} emission by \citet{Veilleux4388}. 
Because of the large FOV
covered by the Fabry-Perot interferometer observations, 
they were able to identify a 
strongly S-shaped form in the isovelocity contours in the galactic disc. 
For the gas velocities that we observe, our limited FOV only allows us 
to identify 
agreement between the general orientations of the central velocity 
features in our data and those of \citet{Veilleux4388}.

We detect two regions of high velocity dispersion, especially 
in [\ion{O}{iii}] maps: 
in the south close to the peak emission, and in the north of the
FOV. These regions      
are characterised by anomalous emission-line profiles, the former
with weak red wings, the latter with a marked additional blueshifted  
component separated by 
$\sim800-900$\,km\,s$^{-1}$ from the main emission component both in 
[\ion{O}{iii}] and {H$\beta$}
\citep[which is significantly higher than reported by][]{Veilleux4388}.  
Examples of spectra and their one-component fits are presented in
Fig.\,\ref{NGC4388maps}, from which it is clear that the velocity maps follow  
the strongest 
component of the lines. The additional redshifted and 
blueshifted components are restricted to the two 
regions and never outshine the main component.

The stellar population modelling shows the largest contribution of young stars
($\sim\!10\%$ of mass) close to the point of inversion of the stellar 
velocities.
The observed field is otherwise dominated by old stars, consistent with the
findings of \citet{Storchi90}.

\subsubsection*{{NGC 5728}}

The host galaxy has a bar along $\mathrm{PA} =33\degr$, a nuclear bar at
$\mathrm{PA} \sim86\degr-90\degr$ \citep{Wozniak95,Emsellem01}, and three rings. 
The nuclear ring has dimensions $9\arcsec\times7\arcsec$, and the 
major axis at $\mathrm{PA} =20\degr\,.$
The Seyfert core is known as one of the most spectacular examples
of the biconical emission-line structure in optical observations, 
with the north-west cone at $\mathrm{PA} =304\degr$ and the south-east cone at
$\mathrm{PA} =118\degr$, and opening angles  $\sim55\degr-65\degr$ 
\citep{Pogge,Wilson93}. 
The emission maxima in both {H$\alpha$}+[\ion{N}{ii}] and [\ion{O}{iii}] 
are located $\sim\!\!1\arcsec$ from the cones' apex \citep{Wilson93}.

The kinematics of {NGC 5728} 
has been characterised by controversial conclusions, 
and so far none of the 
proposed models (radial outflow, inflow or non-axisymmetric rotation in a 
barred potential) has managed to account fully for all the observational 
results. 
The inflow model is supported by the observed asymmetry in the line-emission
distribution and the degree of ionisation 
\citep{Schommer88,Wilson93,Riffel08}: the shorter and less ionised
north-west cone can be explained by obscuration and interaction
with the galactic disc if the north-west is the far side 
(see Figs. \ref{NGC5728maps} and \ref{NGC5728vel} for a better understanding). 
On the other hand, HST observations revealed a sharp apex to the 
north-west cone, whereas the south-east cone terminates with a blunt end
$0.5\arcsec$ from the core, suggesting obscuration by foreground material
\citep{Wilson93}. 
Together with an excellent alignment with the radio emission, this might 
be considered evidence of an outflow \citep{Schommer88}.

Observed with the use of the Fabry-Perot spectrometer by \citet{Schommer88},
strongly S-shaped {H$\alpha$} isovelocity contours in the central 
$10\arcsec$ (2\,kpc) suggested that these features originated 
in the bar potential. On the other hand, the bar is in 
a position for which models by \citet{Roberts79} do not predict velocity field
distortions, therefore \citet{Schommer88} drew no conclusions about the 
importance of the bar. SPH simulations by \citet{Perez5728}, which
included a bar, failed to reproduce the observed velocity field and 
the authors proposed that either an additional nuclear bar or radial flows are
present.

\bigskip

The major axes of the emission structures determined from our OASIS data 
are aligned at $\mathrm{PA} \sim125\degr$ (as measured from the outer contours). 
The emission is asymmetric in the two lobes, as reported before in the 
literature (see above).  
The north-west lobe shows an elongation at $\mathrm{PA} \sim-40\degr$ in 
the inner $4\arcsec$
($\sim\!\!800$\,pc), consistent with the HST observations of \citet{Wilson93}, 
while the more prominent south-west lobe bends towards $\mathrm{PA} 
\sim115\degr.$
We partially resolved the secondary maximum of emission located 
$\sim\!\!2\arcsec$
north-west of the maximum, which was reported e.g., by \citet{Wilson93}.    
The hydrogen emission maps show a prominent arc perpendicular to the
lobes, at a distance of $\sim\!\!5\arcsec (1\,\mathrm{kpc})$  north-west of  
the centre.  We interpret this as part of the ring reported e.g., by 
\citet{Wilson93}.

The gas velocity maps and their one-dimensional cuts (Fig.\,\ref{NGC5728vel}) 
show a strong
gradient of LOS velocities along the major axis of emission, in the vicinity of
the dynamic centre of the galaxy. The velocity maxima of opposite signs 
(approximately $\pm300$\,km\,s$^{-1}$) are separated by only 
400\,pc and are located at the
primary and secondary maxima of emission intensity.   
A comparison with the Fabry-Perot observations \citep{Schommer88} 
provides a different view of the velocity maps across a wider field, providing
thus a larger context which is inaccessible with our
limited FOV: a strong S-shape is found, 
which must be taken into account in the
interpretation (in a future paper). 
The hydrogen velocity field of the outer parts of our FOV (outside the central
$500$\,pc) has an orientation corresponding to the stellar motions that we 
measured.   
We suppose that the interpretation of the kinematic structure of 
{NGC 5728} will be aided by rigorous disentangling of 
multiple-component emission lines that we found to be
distributed within an         
elongated region approximately perpendicular to the
photometric major axis ($\mathrm{PA} \sim45\degr$ in [\ion{O}{iii}] and 
$\mathrm{PA} \sim60\degr$ in {H$\alpha$}) 
and passing through the dynamic centre. Within our single-Gaussian model of 
emission profiles, the spatial region of asymmetric lines is represented by the
large-FWHM area of the map (Fig.\,\ref{NGC5728vel}).    
We illustrate the multi-component emission profiles with the varying 
importance of the individual components in Fig.\,\ref{NGC5728spec}.        

Another interesting result is that for the dust distribution. 
The map of interstellar reddening plotted in Fig.\,\ref{NGC5728maps} shows 
maximum extinction in the regions of emission maxima 
and the star-forming \ion{H}{ii} ring. The obscuration 
of the ionisation cones is also asymmetric and supports the interpretation
of the north-west cone interacting with the galactic disc, and the 
south-east cone situated above the disc.     

The stellar population ages resulting from our SSP modelling show a clear 
systematic distribution, with significant contributions (more than a few 
percent) from young stars only outside the central $\sim\!\!600$\,pc.

\subsubsection*{{NGC 5929}}

{NGC 5929} 
forms an interacting pair with NGC 5930 \citep[e.g.,][]{Bower94}.
The galaxy hosts a linear radio source with two lobes separated by $1\arcsec$
along $\mathrm{PA} =60\degr$ \citep{Whit86,Ulve84},
close to the galaxy major axis at $\mathrm{PA} =45\degr$ \citep{Schm97}.
Optical emission in [\ion{O}{iii}] and H$\alpha$ + [\ion{N}{ii}] 
is in the form of  
cones co-aligned with the radio source, characterised by the opening angle 
of $80\degr$ and an extent of $150-175$\,pc, and 
separated by a perpendicular dust lane \citep{Bower94,Schm96}. 
\citet{Ferr5929} performed    
IFU observations in {H$\alpha$}, [\ion{N}{ii}], and [\ion{S}{ii}] 
with the TIGER spectrograph, and tested several models 
to account for the optical and radio 
emission. They place constraints on the interaction of the radio ejecta with
the ambient gas, i.e.,  
jet/cloud interaction, 
jet/ISM interaction, plasmon expansion, and the bow shock.
None of the models were able to interpret the data satisfactorily.  

\bigskip

Our observations were completed in one spectral domain (MR1) only.
We found differences between the major axes of emission in 
[\ion{O}{iii}] ($\mathrm{PA} \sim90\degr$)  
and in {H$\beta$} ($\mathrm{PA} \sim75\degr$), see Fig.\,\ref{NGC5929maps}. 
A similar discrepancy was seen in the zero-velocity 
curve orientation (approximately perpendicular to the emission major axes) 
in the maps of mean LOS velocities. 
However, the main features of the velocity fields, as well as their 
one-dimensional cuts, were consistent for both ions.  

Our map of stellar velocities showed 
strong misalignment with the gas velocities: the stellar kinematic axis 
($\mathrm{PA} \sim30\degr$) was almost perpendicular to the symmetry axis of 
the gas. 
Moreover, the polarity of the stellar velocity field was reversed 
with respect to gas: in the north-east region, stars had receding 
motion, while the gas was approaching.

The LOS velocity dispersion maps show a stripe of maximum dispersion
at $\mathrm{PA} \sim160\degr$ through the optical nucleus, approximately
perpendicular to the photometric axis of {H$\beta$} (Fig.\,\ref{NGC5929maps}). 
The large-FWHM region is associated with asymmetric line profiles
(Fig.\,\ref{NGC5929maps}), with a separation of 
$\sim\!\!900$\,km\,s$^{-1}$ between the line components.  

The [\ion{O}{iii}]/{H$\beta$} intensity ratio distribution follows 
the morphology of 
[\ion{O}{iii}] emission, with maxima at both the nucleus and off 
the nucleus at $2\arcsec$ east, $2\arcsec$ west, and $4\arcsec$ west. 
The stellar populations show a clear pattern with a growing contribution 
of young populations toward the edges of the observed field.

\section{Discussion \label{discuss}}

Despite the advances in 3D spectroscopic techniques,
optical datacubes obtained with the resolution that we present in this paper
remain rare. 
We have shown the wealth of information that the 
OASIS IFU data can
provide, making it possible to address     
several important problems concerning AGN, 
while opening a perspective of promising
innovative, detailed modelling. 
The relatively large number of emission lines contained in the data not only
allow comparisons of NLR morphologies and kinematics for 
different ionic species,
but also provide diagnostics for probing the physical conditions of
the gas, based on the line intensity ratios. We have been 
able to estimate directly 
from the measured ratios both the electron density 
(the [\ion{S}{ii}] doublet ratio) 
and the interstellar reddening due to dust (the Balmer line ratio). 
The determination of other characteristics such as the ionisation 
parameter, metallicity, or temperature will require future
modelling using spatially resolved diagnostic diagrams presented
in this paper. 

The emission data results have been compared with the stellar data,  
the kinematics in particular. We have also modelled the contributions of
stellar populations of different ages and their variations across the observed
field.   
This simultaneously-acquired, rich information allows a more 
complex understanding of the          
structure and conditions in each of the observed Seyfert galaxies. 
In addition to studying each of the targets individually, 
we take advantage of having a uniform data set of Seyfert 2 galaxies 
and search for general trends in the physical properties. 
In this section, we summarise and discuss the 
obtained results in the context of the entire sample.

\subsection{Non-axisymmetric potentials\label{nonax}}

One of the major goals has been to map the gas velocity fields in detail 
for different emission lines, and compare them with stellar velocities 
if possible, to understand more clearly the emission morphologies, 
the dominant dynamic processes at the galactic centre, and eventually 
the origin of the NLR gas. 
A number of scenarios have been proposed for NLRs that can be tested by 
spectroscopic studies. One of them concerns  
departures from the axial symmetry of the gravitational potential,  
which might be related to the nuclear activity and the gas transport  
across large distances toward the central galactic regions 
\citep[e.g.,][]{Dumas,Schinnerer00}. 
On the other hand, signs of gas motion in 
non-coplanar orbits with the host galaxy may imply an external origin 
of the NLR gas,  
emphasising the role of mergers \citep[e.g.,][]{Morse,Hunt}.  
Streaming motions toward the galactic centre might  
indicate a connection with AGN fuelling, and outflow motions are   
tested for their relation to the radio jets
\citep[e.g.,][]{Bicknell98},  
or to the predicted thermal wind produced by the central torus 
\citep[e.g.,][]{Krolik86,Balsara93}.

Only one of the gas velocity fields obtained from our observations 
seems to be consistent with circular rotation in 
the central kiloparsec -- {NGC~449} 
(Mrk~1) -- as far as our field-of-view allows us to judge. 
The mean LOS velocities of the ionised gas in 
seven of the eleven studied Seyfert\,2 galaxies are characterised by 
S-shaped isovelocity contours in the central regions.
This result supports the findings of \citet{Dumas}, who reported 
velocity contour twists to be more common in the Seyfert galaxies than in 
quiescent ones.  
The S-shapes are in general signatures of  
non-axisymmetric galactic potentials containing bars,    
warps, or nuclear spirals \citep{Roberts79}, and 
were successfully reproduced by these models 
in the case of several Seyfert galaxies 
e.g., by \citet{VeilleuxBar4388,Schinnerer00};    
and \citet{Emsellem06}. 
Their interpretation of the velocity fields by motions in 
non-axisymmetric potentials thus provides an alternative
to the radial flows often invoked to account for deviations from circular
velocities. 
Degeneracies in the interpretation of non-circular
motions were addressed analytically by \citet{Wong},  
who derived prescriptions of how to 
discriminate between bars, warps, and radial flows 
by means of Fourier analysis. 

Beside the S-shaped velocity fields, another type of a non-axisymmetric  
velocity pattern has been represented in our data in the case of 
{NGC~262} (Mrk~348): two striking spots in 
the mean velocity map, suggesting a rotating disc/ring ($\sim\!300$\,pc in 
radius) that is non-coplanar with the host-galaxy disc. 
Concerning outflows, we cannot provide direct evidence, 
even though at least two of the objects in our sample 
({NGC~4388}, {NGC~2992}) 
have had an outflow component confirmed by several authors
\citep[e.g.,][]{Veilleux4388,Marquez98,Allen99,Veilleux2992}. 
We are partially limited by the small FOV, 
but a clearer understanding is expected from our future detailed kinematic
modelling.

Except for one case ({NGC 4388}), the non-circular motions of gas
do not have evident counterparts in the stellar velocity fields.
This is unsurprising since gas, even in the absence of
an AGN-driven  
outflow, is in general not expected
to follow the same orbits as stars, due to its dissipative nature
resulting in lower velocity dispersion, its higher responsiveness
to non-axisymmetric perturbations, possible ring formation, and a
phase shift with respect to axes of stellar orbital families.
A counter-rotation or tilt with respect to the plane of the stellar disc
are also plausible in the case of gas having an 
external origin (e.g., as a result of a minor merger).
Observed misalignments between the kinematic axes of the stellar and gaseous
components in active and non-active galaxies were discussed  
in detail by \citet{Dumas}, who concluded that departures are both more
frequent and significant in galaxies hosting an AGN. They also found that 
the closely aligned axes and a lack of isovelocity contour twists 
in gas were correlated with low-accretion rate AGN. Their results  
suggested a close relation between the AGN and the dynamics of the host.

\subsection{Nuclear rings} 

Further type of evidence of a non-axisymmetric potential is the presence of
nuclear rings. A small fraction of observed rings may be due to collisions or
mergers of galaxies \citep[e.g.,][]{Knapen04}, or to accretion of 
intergalactic gas, although the vast majority
are probably resonance phenomena \citep{Buta96}. Nuclear rings have been
identified in approximately 20\% of spiral galaxies \citep{Knapen05}.
Our sample of Seyfert\,2 galaxies possesses at least one object with a 
nuclear ring, reported previously \citep{Wilson93} and 
confirmed in our data: {NGC~5728}.   
We detect the ring in the surface brightness maps of the Balmer lines
and low-ionisation lines of metals, while it is missing in [\ion{O}{iii}]. 
We have found it to be associated with high interstellar extinction, 
which suggests that it contains large amounts of dust.

Evidence of other
rings might originate in the kinematic measurements, as was the case for 
NGC~5252 in \citet{Morse}, and will be tested in 
future modelling. 
The strongest candidate to contain a ring is {NGC~262} (Mrk~348) 
with the striking velocity pattern described
in Sect.\,\ref{nonax}, not reported so far elsewhere.  
As suggested by \citet{MulObs},
{NGC 262} possibly hosts another ring at $\sim\!\!1$\,kpc.
Marginal evidence of its detection was also 
found in our data, in the mean LOS velocity map of [\ion{O}{iii}], 
which displays  
systematic velocity pattern at the edges of our FOV that is  
qualitatively consistent 
with rotation in an opposite sense to that of the nuclear ring.

\subsection{NLR morphologies, cones}

The morphologies that we have obtained from analysis of different 
individual lines 
are not always identical or even aligned along the same axis. 
The misalignments between emission in [\ion{O}{iii}] and 
low-ionisation lines 
(e.g., {NGC~449}, {NGC~5929}) 
provide evidence of additional dynamic processes  
affecting the high-ionisation components of the gas, such as the radio jets
that accelerate and/or ionise the gas.   

Two of the galaxies ({Mrk 1066}, 
{NGC 5728}) show prominent regions detectable only 
in hydrogen and low-ionisation lines, which are virtually missing 
in [\ion{O}{iii}]: a nuclear ring in {NGC~5728} 
\citep[reported before e.g., by][]{Wilson93},  
and an off-centre region of diameter $\sim\!500$\,pc in 
{Mrk~1066}. The regions are 
characterised by a ratio [\ion{O}{iii}]/{H$\beta$} markedly 
lower than in the remainder of the 
observed field, and have therefore probably not been ionised by the AGN 
radiation.  
They are also associated with high interstellar 
reddening, which indicates a high amount of dust and possibly  
star formation, and hence an ionisation by young massive stars. 
Marginal support of this interpretation comes from the maps of stellar ages,
which show relatively large contributions from young stars in the specified 
locations compared to the neighbouring regions. 
Ionisation by sources other than the active nucleus thus 
definitely play a role in the observed NLR morphology.

Directly related to the Unified Model of 
AGN (and the existence of the nuclear torus), the major puzzles about NLRs 
are their extent, the 
existence of ionisation cones, and the intrinsic geometry of the gas 
distribution. Significant evidence of cones was found in 
ground-based and HST images \citep[e.g.,][]{Wilson93,Falcke98}. 
The NLR structure parameters were analysed statistically for large 
samples by \citet{MulObs}; and \citet{Schm03Obs}, who both concluded that
a surprisingly low number (with respect to the predictions of the 
Unified Model) of NLRs in Seyfert\,2 nuclei 
have a major axis that is significantly longer than their minor axis. 
We reached the same conclusion, with the caveat that our results are 
influenced by both the small FOV and spatial resolution, mostly determined    
by the seeing conditions ($\sim\!\!1\arcsec$). 

\citet{MulSey} made predictions of NLR extents and morphologies 
for different observer viewing angles, and also for  
different intrinsic gas distributions ionised by the conical radiation 
field from the AGN. They considered two basic geometries of the gaseous 
region, a sphere and a disc, and tested their model predictions 
with data from a large  
imaging survey. In this context, our kinematic measurements clearly 
show that the observed NLRs are characterised by mostly rotational 
(although probably non-circular) motions, and are thus non-spherical.

\subsection{Alignment with radio emission}

Where detected, the elongated optical emission structures have been found to be 
aligned with radio emission data published in the literature, 
to within the measurement errors. This alignment infers an alignment 
between the torus and accretion-disc axes 
and/or an influence of the radio jet on the NLR gas dynamics and ionisation
\citep{Schm03Results}.
Our result confirms previous studies of the alignment 
between the two wavelength bands, demonstrated e.g., by    
\citet{WilsonOrient,Nagar99Orient,Falcke98,CapettiRadio}; 
and \citet{Schm03Results}.

As discussed above, {NGC 449} and {NGC 5929} 
of our Seyfert sample have different 
orientations  
of photometric axes in the high-ionisation and low-ionisation emission lines.  
The radio-jet axis is
consistent with [\ion{O}{iii}] emission in {NGC 449}, and with 
{H$\beta$} emission in {NGC 5929}. 
Conclusions are ambiguous for {NGC 2273}, 
which exhibits different shapes in its radio continuum
emission at 20\,cm and 6\,cm wavelengths \citep{Ulve84}. 
The {H$\beta$} emission appears to agree with the former 
and [\ion{O}{iii}] with the latter. 

\subsection{Emission-line profiles}

We detect prominent splitting of emission lines in eight of the observed
NLRs, usually limited to only part of the FOV, identified as a large-FWHM 
region in the maps produced by our single-Gaussian line-profile modelling. 
In seven of the galaxies the region is of 
elongated shape approximately perpendicular to the NLR major axis 
and passes through the dynamic centre. Multi-component emission-line
profiles
have been reported in numerous NLR spectroscopic studies, but the
integral-field spectroscopy permits us to map in detail the spatial variations
in each component and to use the information in modelling the 
NLR structure
and kinematics. The origin of the line splitting found in our data is unclear
and will be addressed in future modelling of the observed
objects.    
The most common interpretations include superposition of several
kinematic components \citep[e.g.,][]{Schommer88}, or expanding outflows 
\citep[e.g.,][]{Capetti99}, rapidly rotating circumnuclear discs
\citep[e.g.,][]{Rubin4388}, and also bar-induced non-circular motions 
\citep[e.g.,][]{Veilleux2992,Bureau99,Kuijken95}.

\subsection{Electron densities}
We have estimated electron densities from the [\ion{S}{ii}] doublet in 
six galaxies of our sample  
for which data are available. In all cases, a non-uniform density
distribution with a central peak has been found,  
the peak values ranging between $\sim\!600$\,cm$^{-3}$~and 
$\sim\!\!1500$\,cm$^{-3}$, and 
declining to $\sim\!200$\,cm$^{-3}$~outside the central $\sim\!500$\,pc.    
The derived inhomogeneous gas distributions may have a strong effect on the 
estimates of global quantities, such as the NLR mass and the Str\"omgren
radius. 

The distributions are asymmetric and the aforementioned values are thus
direction-dependent, although the overall results are consistent with 
long-slit observations \citep[e.g.,][]{BennertSey2,Fraq}.    
More precise modelling is necessary 
to decide which of the values in the outer parts of the FOV (which are often
relatively high) estimate reliably the intrinsic densities and which 
are an effect of measurement errors, to which the derived density values are
highly sensitive.                  
We also note that both 
the density values and the radial profiles are influenced by the 
method of derivation. The [\ion{S}{ii}] ratio is density-sensitive 
only within a limited 
range of values, and provides a density measure only in regions where these
sulphur ions are present.
The effects of observational errors are also non-negligible.  
We have also derived the densities based on the
assumption of an electron temperature of $T=10\,000$\,K, and 
different temperatures in different regions may therefore modify the absolute 
values and the spatial distribution of the computed densities 
\citep[see e.g.,][]{Osterbrock89}.

\subsection{Interstellar reddening}

The dust distribution may contribute significantly to the reconstruction 
of the galactic structure and the orientation of the observed nuclear
emission regions with respect to the host galaxy, such as emission cones 
above or underneath the galactic disc. 
We have been able to estimate the dust distribution from the ratio of Balmer
lines in 7 objects. The maximum extinction is usually connected with regions
of maximum emission. Distinct off-nuclear regions of high extinction have been 
found in {NGC~262} (Mrk~348) and {NGC~5728}, 
consistent with low-ionisation regions
such as the star-forming ring in {NGC~5728}.       
The dust distribution in 
{NGC~5728} is an excellent example of asymmetry: the obscuration is
much higher in the north-west emission cone than in the south-east part 
(see Fig.\,\ref{NGC5728maps}), which is consistent with the interpretation
of \citet{Schommer88}, and \citet{Wilson93}, who predict the north-east cone 
to be interacting with the galactic disc.

The values of $E(B-V)$ that we have derived were based on the
assumption that the intrinsic {H$\alpha$}/{H$\beta$} 
ratio value equals 3.1, as is
standard for AGN-excited gas due to hard radiation causing a relatively 
large transition
zone \citep{Osterbrock89}. However, not all of the observed
spectra necessarily originate in AGN-excited locations with an overabundance
of high-energy photons compared with stellar spectra, and the intrinsic value
of 2.86 would have been more appropriate. Therefore, reddening is
underestimated in these locations.

The $E(B-V)$ that we have obtained is strongly dependent on the stellar 
population model applied to correct the Balmer emission lines for 
underlying absorption. 
The computed $E(B-V)$ values are significantly higher than 
those published e.g., by \citet{BennertSey2,Raimann03,Fraq}; and 
\citet{HoIII}.  The values that are common in the literature 
are of below one magnitude, in many cases determined  
by methods that differ from that of the Balmer decrement, 
and are thus independent of the stellar absorption models.  
We have tested the influence of the SSP models on the 
spatial distribution of dust, and on the $E(B-V)$ values. We have found 
that the spatial distributions remain unchanged qualitatively 
after applying the ``minimum absorption correction'' and using 
an old SSP as a stellar template. In contrast, the distribution differed 
when we did not correct for absorption and subtracted only 
the smooth continuum. 
We considered the no-absorption case to be
non-realistic, and therefore conclude that 
our maps of $E(B-V)$ reflect the true character of the dust distribution. 
At the same time, the problem of the high reddening values was not solved by 
applying a different stellar model.

\subsection{Diagnostic line ratios}

The maps of diagnostic line ratios trace the spatial 
distribution of the gas physical characteristics 
and the properties of the ionisation source.  
The [\ion{O}{iii}]/{H$\beta$} ratio provides the 
most straightforward interpretation by tracing   
the gaseous media ionised by
different sources projected along the line of sight. An outstanding 
example is {NGC 4388} in \citet{Veilleux4388}, where 
[\ion{O}{iii}]/{H$\beta$} helps to separate between the
gas in the galactic disc and the extraplanar material. An impact of the 
ionisation maps has been demonstrated in the case of {NGC 5728}, 
where the higher
ionisation in one of the lobes has been used to advocate the inflow
interpretation of the observed velocity field due to the position of the 
ionisation cones in relation to the galactic disc \citep{Schommer88}.

In the objects that we have studied, 
maximum ionisation occurs in the vicinity of the galactic nucleus, and 
is consistent
with a model of ionisation by AGN radiation.    
The isocontours traced by the [\ion{O}{iii}]/{H$\beta$} are not 
identical to those of the [\ion{O}{iii}]
emission, and in approximately half of the sample the symmetry changes,
which implies a contamination by other emission sources. The ionisation 
maps in general show more clearly defined axes than the 
[\ion{O}{iii}] emission, and have a more conical form than 
the [\ion{O}{iii}] maps, as far as 
our spatial resolution allows us to perceive.  
As already discussed in previous subsections, the ionisation maps have helped
us 
to identify a ring in {NGC~5728} and probably a star-forming region in 
{Mrk~1066}. We have also confirmed 
high ionisation in the southern lobe of {NGC~5728}, reported by
\citet{Schommer88}. The [\ion{O}{i}]/{H$\alpha$}, 
[\ion{N}{ii}]/{H$\alpha$}, and [\ion{S}{ii}]/{H$\alpha$} 
ratios have distributions that are off-centre in most of the observed cases. 

We have constructed a set of three spatially resolved diagnostic diagrams for 
7 of the 11 observed Seyfert 2s (for {Mrk~34} 
only two have been possible).
The distributions of points in the diagrams show considerable differences 
between the NLRs, 
reflecting important differences in their inner structures. 
We have included several theoretical curves 
\citep{Kew01,Kauf,StasiAGN,Kew06} classifying emission-line 
objects based on their photoionisation source. However, the division inside the
plots is only informative, since the models correspond to spectra
integrated spatially over the nebula and are not adapted to the spatially
resolved diagrams. A numerical ionisation modelling is necessary to achieve a
correct interpretation. 
A strong Seyfert regime across the entire observed range of projected radial
distances is detected in {Mrk 34} and {NGC 262} (Mrk 348). 
The opposite is true for {Mrk 622},
where all of the points lie below the Seyfert/LINER limit (which may be a
consequence of the applied SSP models). The distributions of points are 
much less
compact for {Mrk 1066}, {NGC 449}, {NGC 2992}, and 
{NGC 5728}, which have regions
in the Seyfert regime as well as regions consistent with non-Seyfert   
ionisation.  
Numerical modelling similar to that carried out by \citet{BennertSey2} will be
necessary to determine the true edges of the NLRs.

\subsection{Stellar populations}

Stellar population modelling performed by using synthetic
young and old SSPs has revealed varying contributions from 
young stars and their
spatial distribution. The occurrence of young stars confirms 
results achieved in previous studies by several independent methods:  
wind lines from O and B stars in ultraviolet (UV) spectra 
\citep[e.g.,][]{Heckman97,Gonz98}, and  
absorption lines of helium and of hydrogen in 
high-order Balmer series from O, B, and Wolf-Rayet stars
\citep[e.g.,][]{Gonz98,Gonz01}.  
Statistical studies of Seyfert galaxies
\citep[e.g.,][]{Stor00,Joguet01,Raimann03} reported 
signatures of young populations 
in $30-50\%$ of the observed Seyfert\,2 galaxies, and significantly higher
contributions ($\sim\!\!20\%$ in terms of 
flux in the nuclei) than in non-active galaxies.  
The detection of recent starburst in the nuclear regions of active galaxies
may be closely related to the interplay between the host galaxy and the AGN, 
and to the triggering of the AGN . 
The highest contributions from young stars are present
in AGN with high IR luminosity, which are interacting and have distorted
inner morphologies \citep{Cid01Starburst,Stor01}.

We have compared the relative mass fractions of young stars 
obtained in our study with the results of \citet{Cid04}. 
The comparison has been completed for 
spatially unresolved data, where we have simulated a $1\arcsec$ aperture 
to comply with \citet{Cid04} (see Table\,\ref{tabyoung}). 
The mass fractions that we obtain for the young population are of the same
order as the combined fractions of young ($<10^2$\,Myr) and 
intermediate ($10^2\,$Myr -- 1.4\,Gyr) populations inferred by
\citet{Cid04}.

\begin{table}[hbt]
\caption{\label{tabyoung}\small
Weights of stellar populations obtained by modelling of
OASIS data, and compared to \citet{Cid04}, on aperture $\sim\!\!1\arcsec\,.$
The symbols $\mu_\mathrm{Y},\mu_\mathrm{I},\mu_\mathrm{O}$ denote mass
fractions
of young, intermediate and old populations, respectively. Analogously for
$\mu_\mathrm{young},~\mu_\mathrm{old}$ in our OASIS data.
}
\begin{center}
\begin{tabular}{|l||ll||lll|}
\hline
Galaxy & \multicolumn{2}{c||}{OASIS data}& \multicolumn{3}{c|}{\citet{Cid04}}
\\
\cline{2-6}
& $\mu_\mathrm{young}$ & $\mu_\mathrm{old}$ & $\mu_\mathrm{Y}$ &
 $\mu_\mathrm{I}$ & $\mu_\mathrm{O}$ \\
\hline
\hline
NGC 2992 & 0.09 &  0.91 &      0.0000 & 0.0260  & 0.9740 \\
NGC 3081 & 0.008&  0.992&      0.0000 & 0.0000  & 1.0    \\
NGC 4388 & 0.02 & 0.98  &      0.0002 & 0.0000  & 0.9998 \\
NGC 5728 & 0.06 & 0.94  &      0.0002 & 0.0699  & 0.9299 \\
\hline
\end{tabular}
\end{center}
\end{table}

At least half of our Seyfert\,2 sample exhibit clear spatial gradients in 
stellar population ages.
In four galaxies ({NGC 3081}, {NGC 4388}, 
{NGC 5728}, {NGC 5929}),
old stars have been found to be dominant
close to the dynamic centre or the location of peak emission, 
and a gradual increase in the
contribution from young stars was measured 
outwards, which contradicts 
\citet{Raimann03} who found an opposite trend. 
On the other hand,
{NGC 2992} contains its youngest populations close to the dust
lane, and {NGC 2273} has a maximal contribution from the young 
SSP north of the
dynamic centre, which gradually decreases outwards. The remaining five 
galaxies ({NGC 449}, {Mrk 34}, {NGC 262}, 
{Mrk 622}, {Mrk 1066}) do not have high-quality stellar data, 
but the 
youngest populations have been fitted visibly in the central-most regions.

\section{Summary and conclusions}

We have analysed a set of 11 nearby Seyfert galaxies of type 2, observed 
with the integral-field spectrograph OASIS, which mapped 
the central $\sim\!\!1$\,kpc 
with medium spatial and spectral resolution. The central goal
of our study was to obtain -- from optical emission-line spectra --
spatially resolved spectral diagnostics, and kinematical maps  
of the ionised gas. 
We have constructed, where possible:\\

\begin{itemize} 
\item
2D maps of: surface brightness in individual lines, line ratios, 
velocity fields (mean LOS velocity and velocity dispersion), electron density,
interstellar reddening, stellar mean LOS velocities, and 
relative mass fractions of young stellar populations;\\

\item
Spectral-diagnostic diagrams for line ratios sensitive to the
source of ionisation (Seyfert, LINER and starburst regimes). \\
\end{itemize}

The 3D technique that we have applied provides a wealth of information
that will be investigated in depth in forthcoming papers.        
The combined advantage of the relatively high spatial resolution and the        
length of the spectral range produce unprecedented details about the
morphology, and both 
the kinematic and ionisation structure of the ionised gas of NLRs. 
We draw the following conclusions from the present study:

\begin{itemize}

\item
Eighty percent of the observed galaxies exhibit S-shaped isovelocity contours 
in the mean LOS velocity maps of the ionised NLR gas. This suggests that  
non-axisymmetric potentials are probably common within the studied 
sample. 

\item
We have detected a possible nuclear ring or a radial flow of gas 
in {NGC 262} (Mrk 348), which was not reported before.
It has been identified based on the mean LOS velocity maps derived 
from the observed emission lines. 

\item
Electron densities usually reach their maximum values 
of $\sim 1000$\,cm$^{-3}$ at the galactic nucleus, and decrease outwards.   

\item
Dust is distributed unevenly in the central regions 
and will have to be included in a future 
detailed study of conditions in NLRs.

\item
Emission-line splitting has been identified in 8 out of 11 objects, 
confined to a clearly defined spatial region at the galactic centre, 
and in most cases elongated 
perpendicular to the major axis of emission.        

\item
Our spatially resolved diagnostic diagrams show radial evolution 
from the Seyfert-like regime in the central regions to other regimes at larger
distances.  

\item
We find spatial gradients in stellar population ages, and confirm the presence
of young stars. 

\end{itemize}

\begin{acknowledgements}
We would like to thank the anonymous referee for comments and
suggestions.    
The work has been supported 
by the Centre for Theoretical Astrophysics (LC06014) of the Ministry of 
Education of the Czech Republic, 
by the Institutional Research Plan No.
AV0Z10030501 of the Academy of Sciences of the Czech Republic,
by the grant No. 205/03/H144/2003/02 of the Czech Science Foundation
and by a PhD fellowship provided by the French government for I.S.
The research has made use of the NASA ADS database,  
HyperLeda database (http://leda.univ-lyon1.fr), and  
NASA/IPAC Extragalactic Database (NED)
which is operated by the Jet Propulsion Laboratory, 
California Institute of
Technology, under contract with the National Aeronautics and Space
Administration. 
{We have made use of software packages for penalised pixel fitting 
and Voronoi tessellation developed by the SAURON consortium.} 
Reduction software packages of ESO MIDAS (version 04SEPpl1.0) and IRAF 
have been used. IRAF is distributed by the National 
Optical Astronomy Observatories, which are operated by the Association of 
Universities for Research in Astronomy, Inc., under cooperative
agreement with the National Science Foundation.
\end{acknowledgements} 

\bibliographystyle{aa}
\bibliography{ref}

\vskip3cm

\begin{figure*}
\begin{flushleft}
  \includegraphics{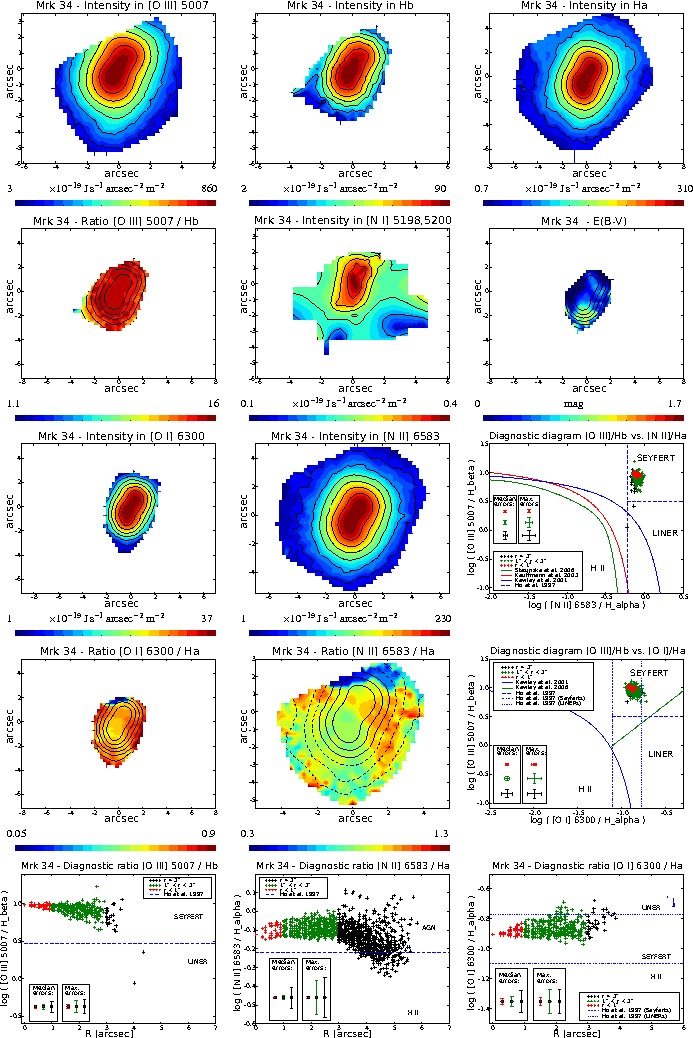}
\caption{\label{Mrk34maps} Mrk 34. See detailed caption
before this graphic section. Surface brightness is in
$10^{-19}$\,J\,s$^{-1}$\,arcsec$^{-2}$\,m$^{-2}$, extinction in mag in
all figures.
}
\end{flushleft}
\end{figure*}

\begin{figure*}
\begin{flushleft}
  \includegraphics{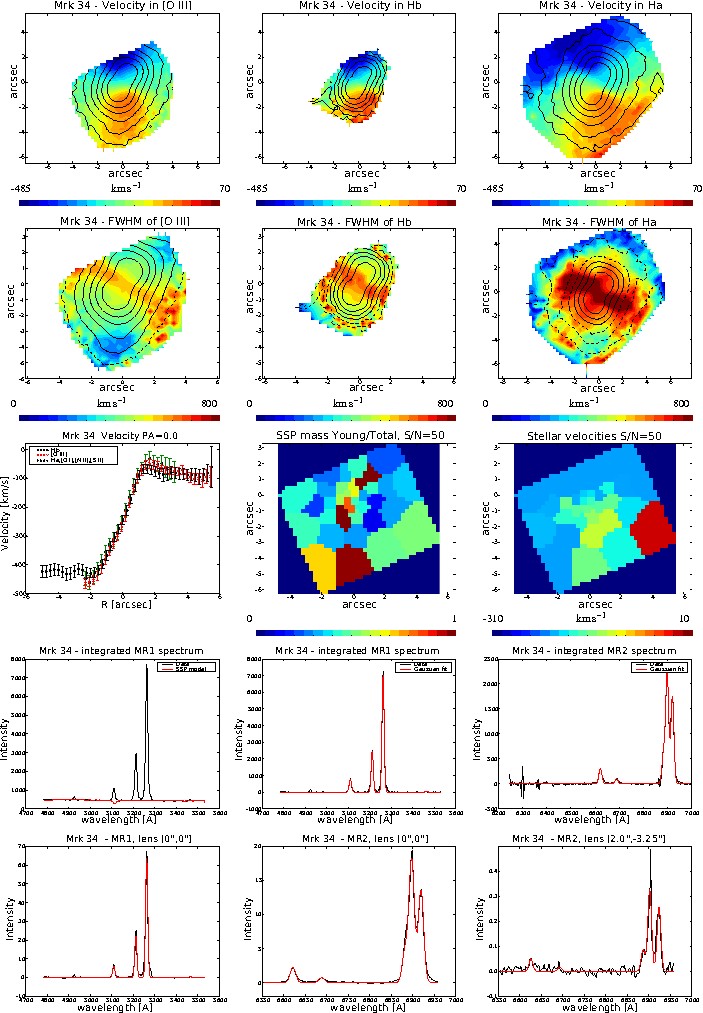}
\caption{\label{Mrk34vel} Mrk 34. {Mean LOS velocities and FWHM are in
unites of km\,s$^{-1}$ in all subsequent figures.}
}
\end{flushleft}
\end{figure*}

\begin{figure*}
\begin{flushleft}
  \includegraphics{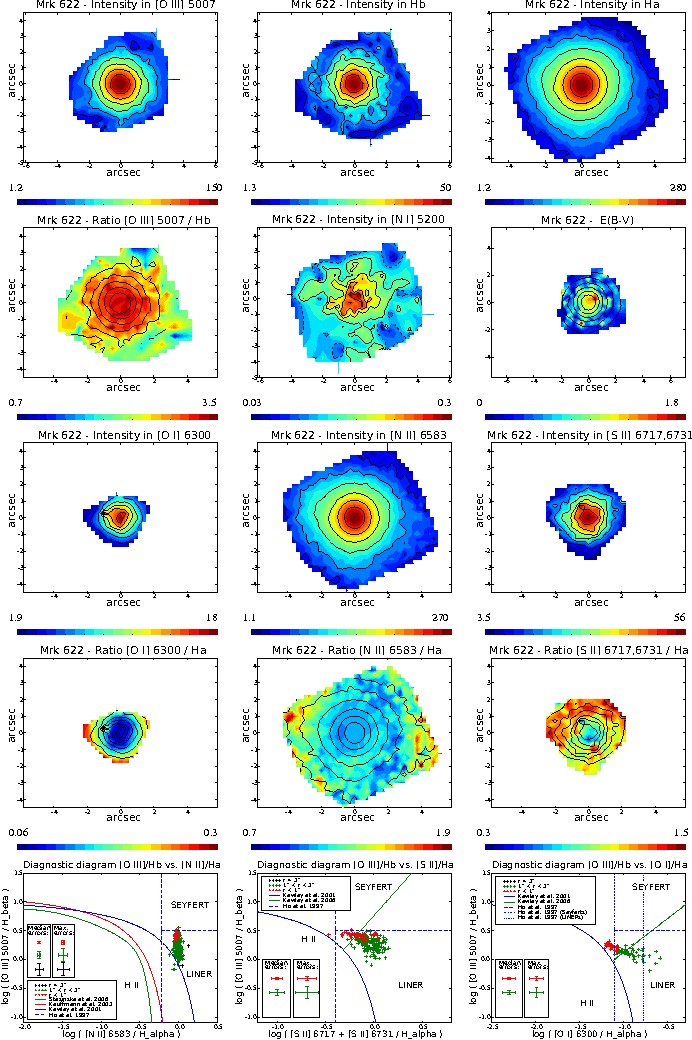}
\caption{\label{Mrk622maps} Mrk 622.
}
\end{flushleft}
\end{figure*}

\begin{figure*}
\begin{flushleft}
  \includegraphics{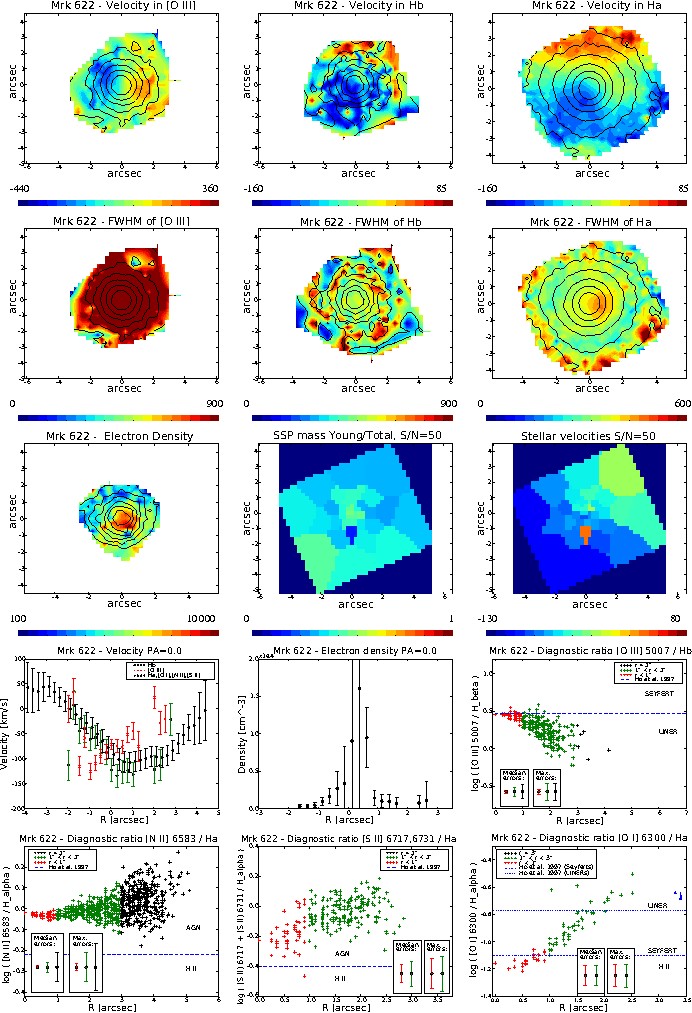}
\caption{\label{Mrk622vel} Mrk 622.
}
\end{flushleft}
\end{figure*}

\begin{figure*}
\begin{flushleft}
  \includegraphics{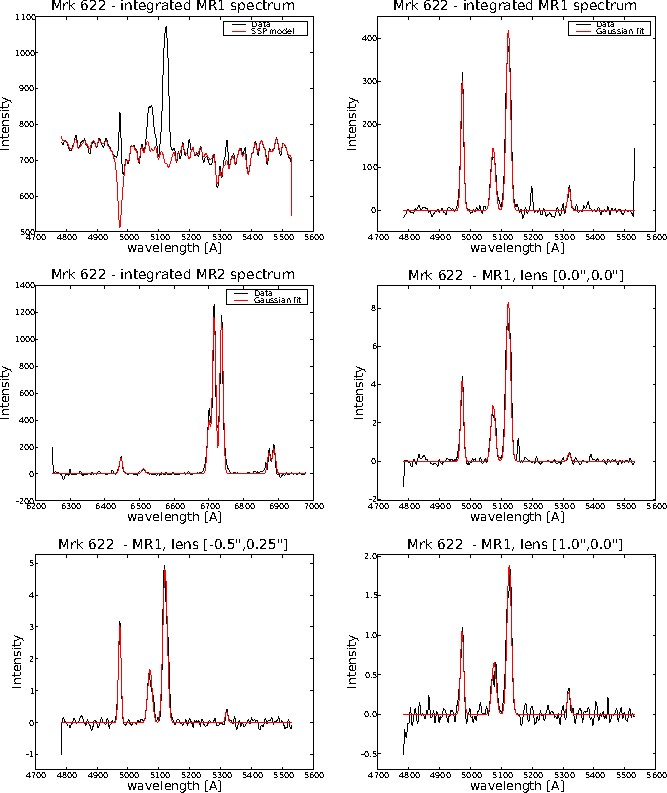}
\caption{\label{Mrk622spec} Mrk 622.
}
\end{flushleft}
\end{figure*}

\begin{figure*}
\begin{flushleft}
  \includegraphics{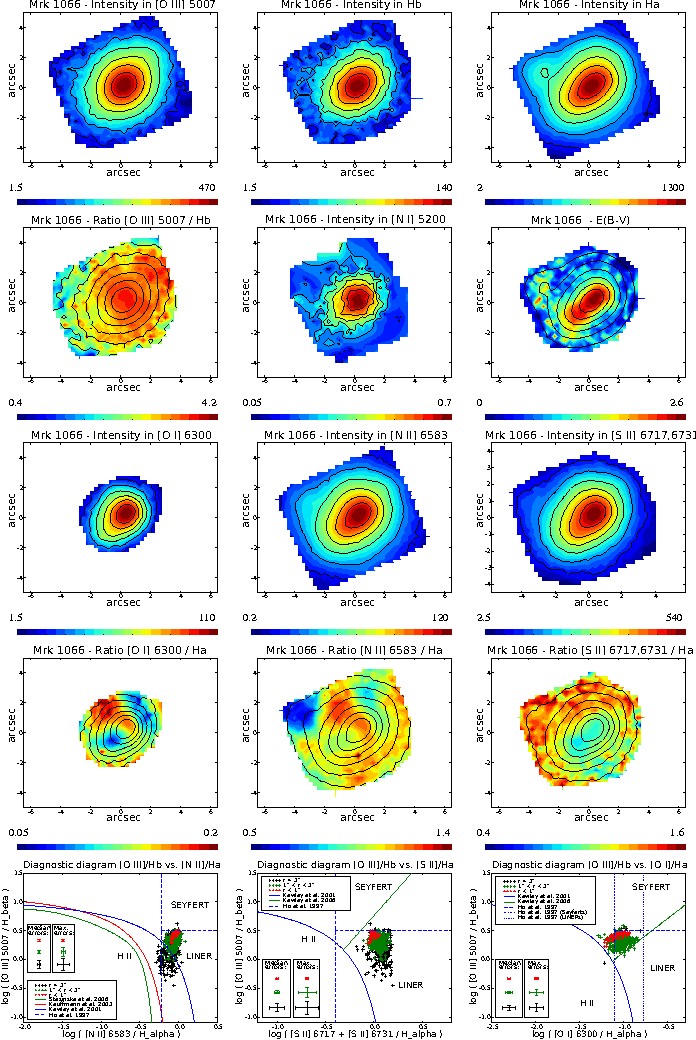}
\caption{\label{Mrk1066maps} Mrk 1066. 
}
\end{flushleft}
\end{figure*}

\begin{figure*}
\begin{flushleft}
  \includegraphics{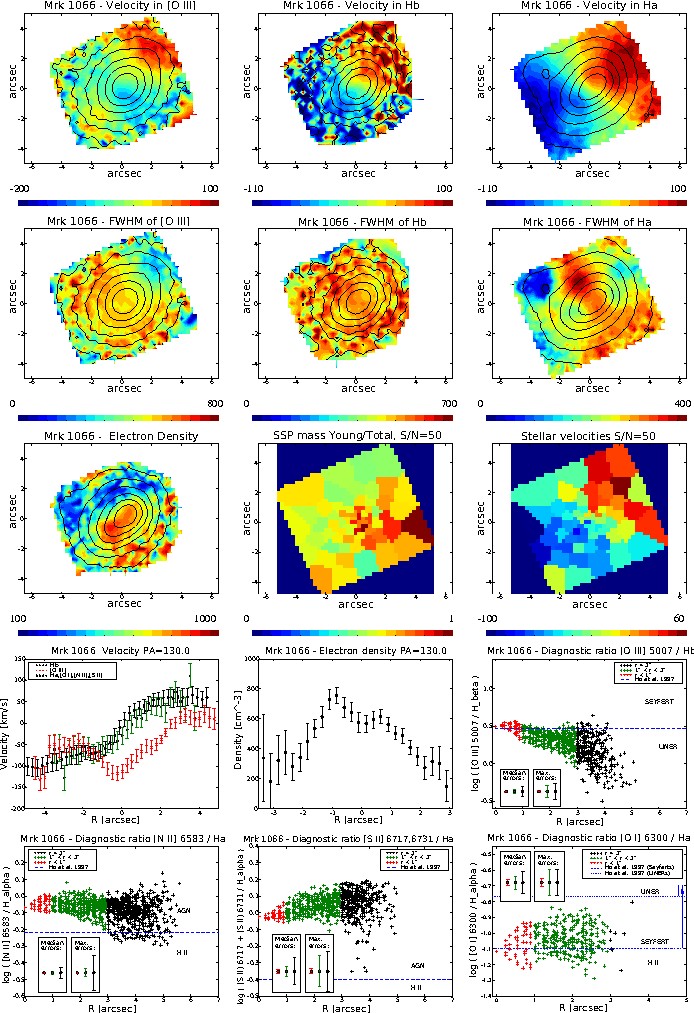}
\caption{\label{Mrk1066vel} Mrk 1066.
}
\end{flushleft}
\end{figure*}

\begin{figure*}
\begin{flushleft}
  \includegraphics{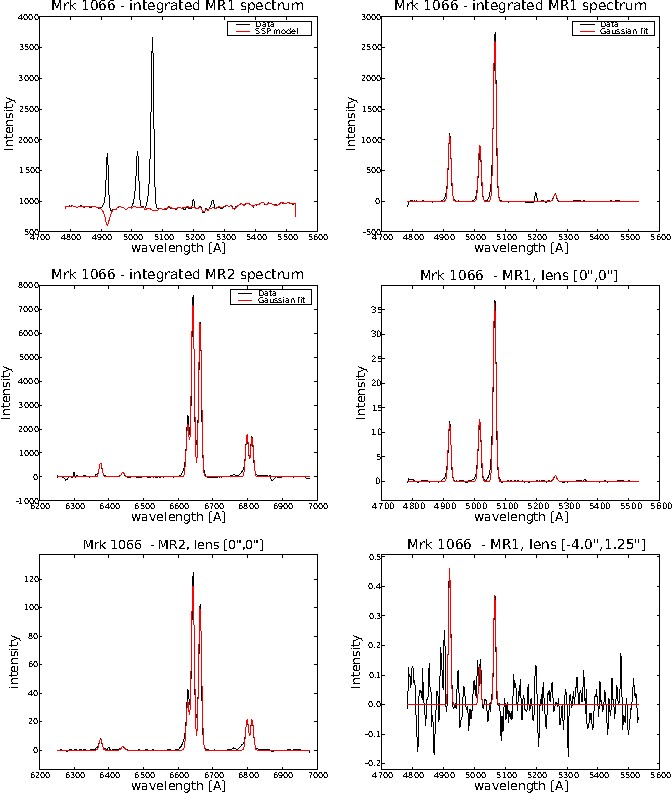}
\caption{\label{Mrk1066spec} Mrk 1066.
}
\end{flushleft}
\end{figure*}

\begin{figure*}
\begin{flushleft}
  \includegraphics{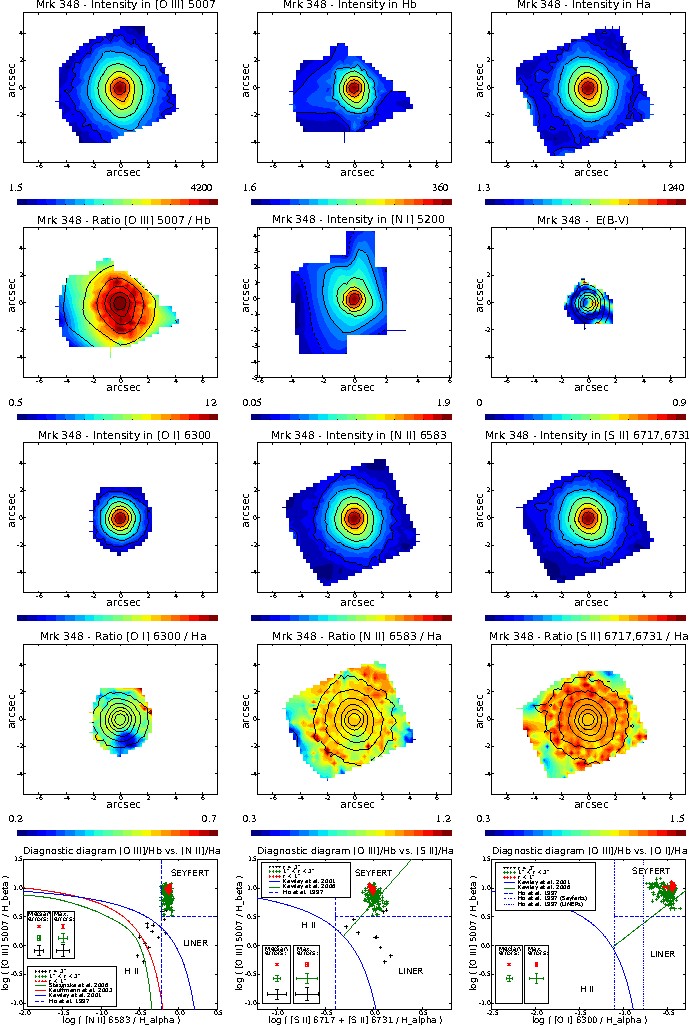}
\caption{\label{Mrk348maps} NGC 262 (Mrk 348). 
}
\end{flushleft}
\end{figure*}

\begin{figure*}
\begin{flushleft}
  \includegraphics{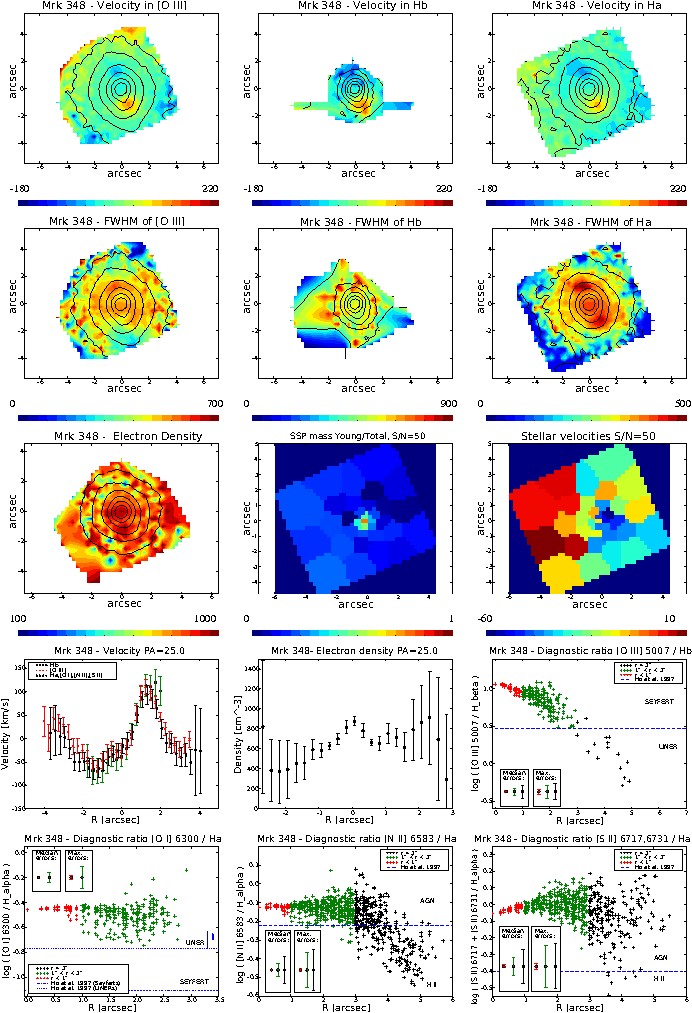}
\caption{\label{Mrk348vel} NGC 262 (Mrk 348).
}
\end{flushleft}
\end{figure*}

\begin{figure*}
\begin{flushleft}
  \includegraphics{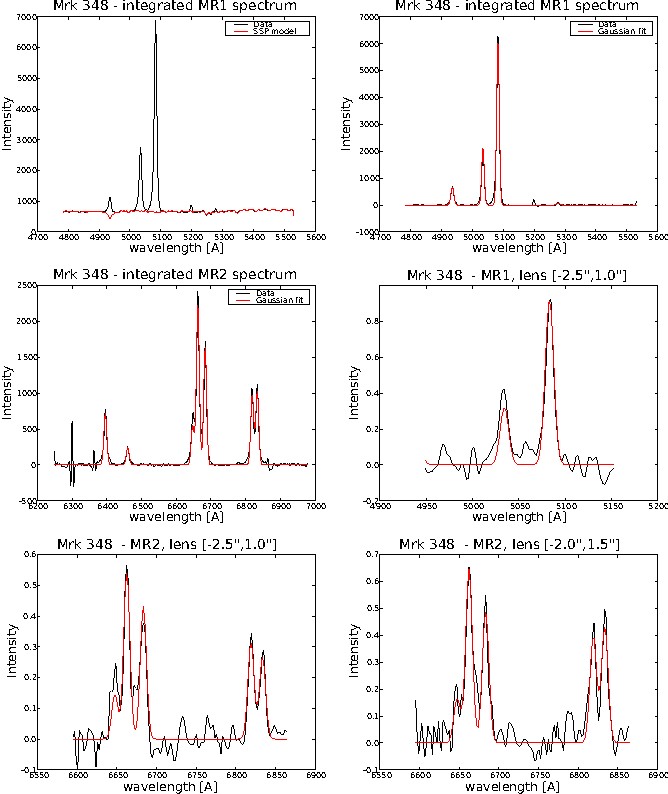}
\caption{\label{Mrk348spec} NGC 262 (Mrk 348).
}
\end{flushleft}
\end{figure*}

\begin{figure*}
\begin{flushleft}
  \includegraphics{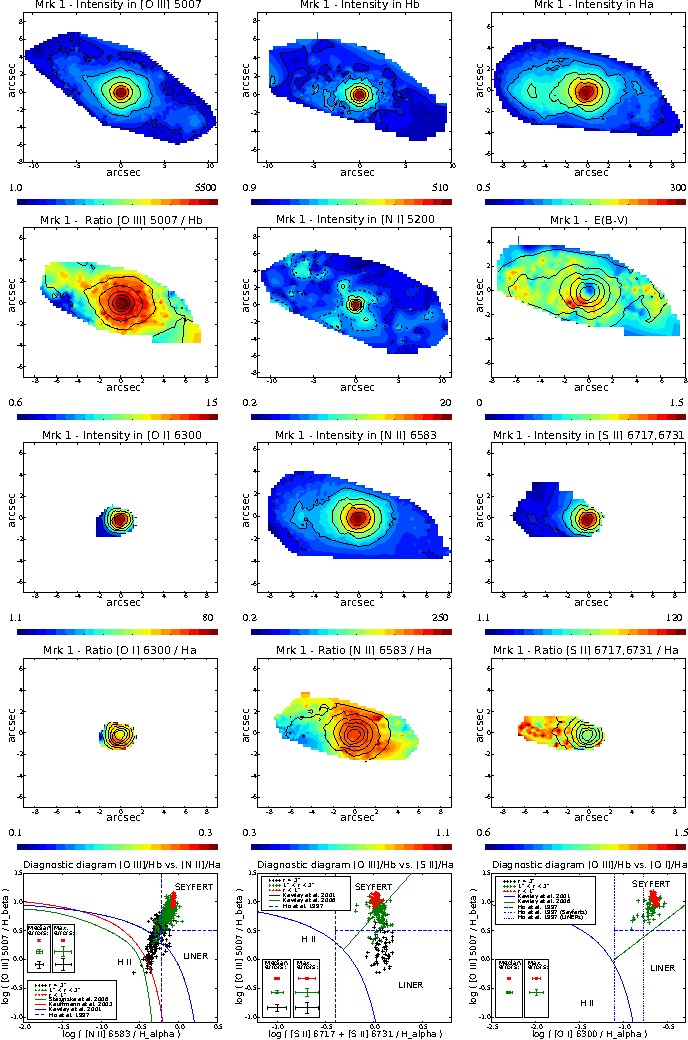}
\caption{\label{Mrk1maps} NGC 449 (Mrk 1).
}
\end{flushleft}
\end{figure*}

\begin{figure*}
\begin{flushleft}
  \includegraphics{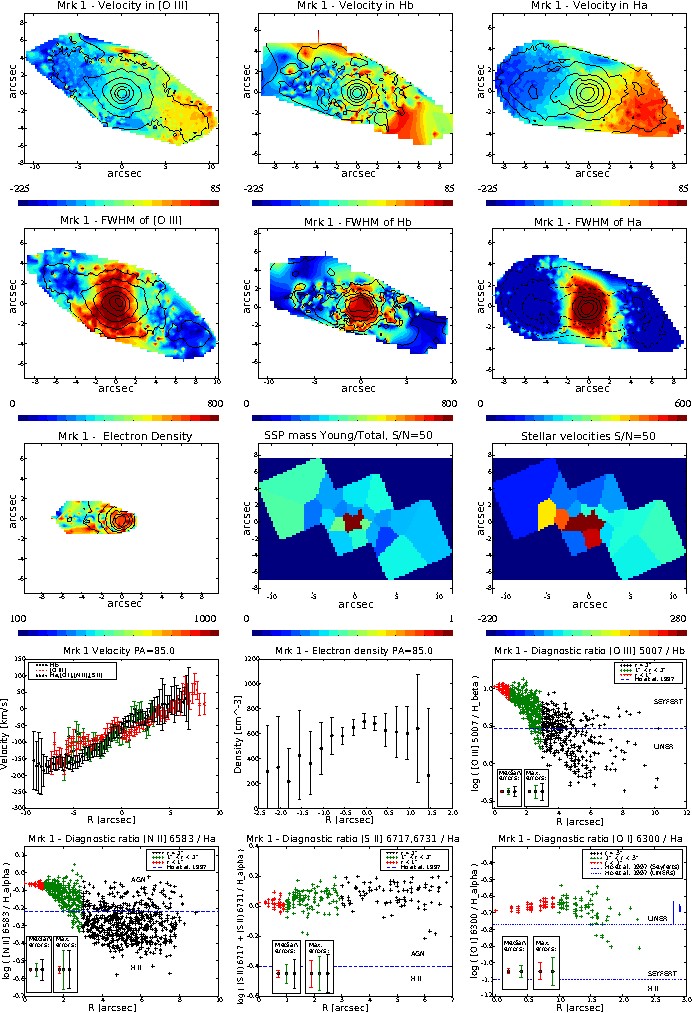}
\caption{\label{Mrk1vel} NGC 449 (Mrk 1).
}
\end{flushleft}
\end{figure*}

\begin{figure*}
\begin{flushleft}
  \includegraphics{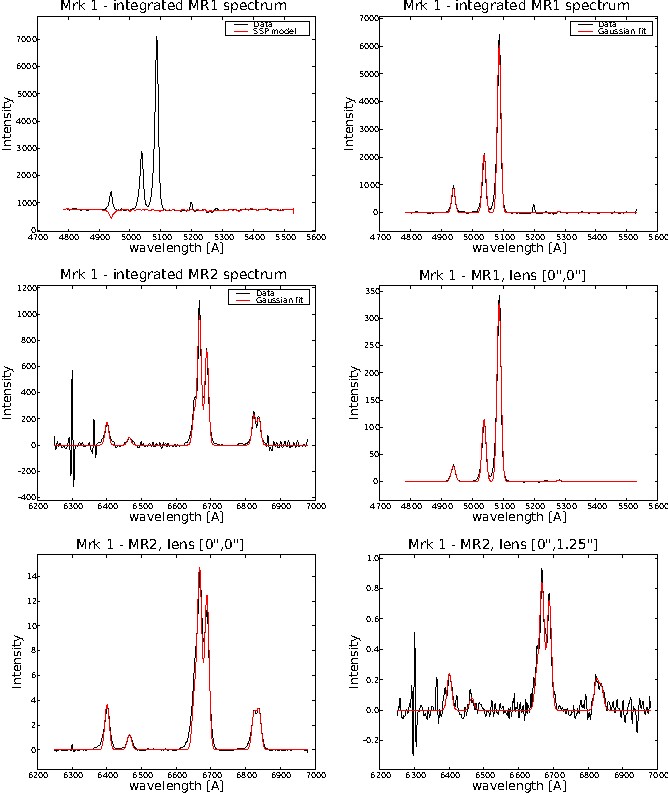}
\caption{\label{Mrk1spec} NGC 449 (Mrk 1).
}
\end{flushleft}
\end{figure*}

\begin{figure*}
\begin{flushleft}
  \includegraphics{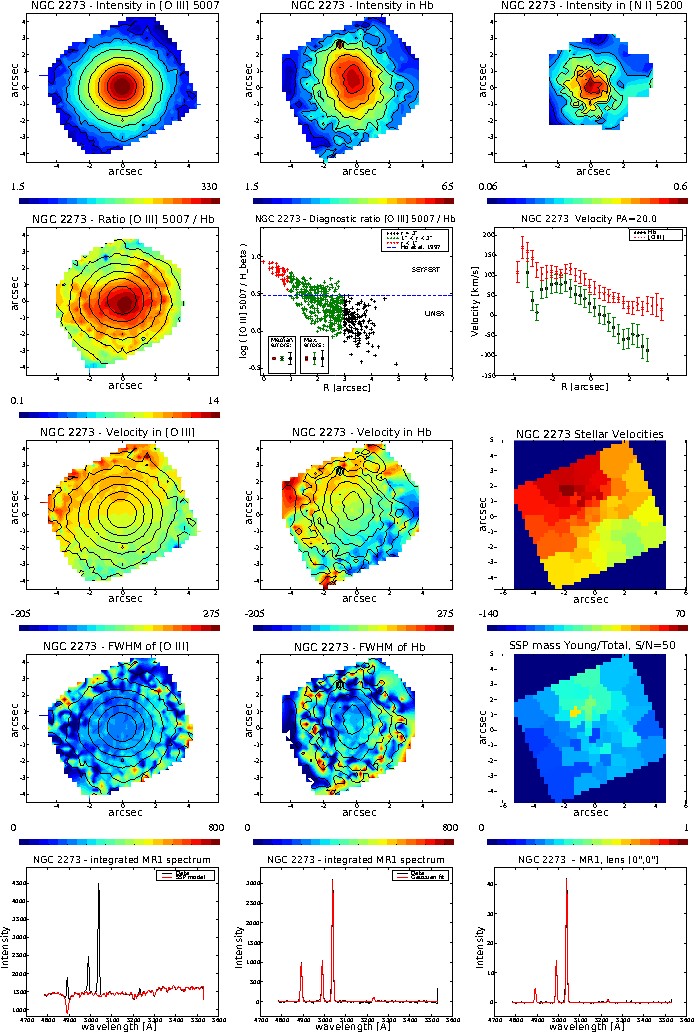}
\caption{\label{NGC2273maps} NGC 2273.
}
\end{flushleft}
\end{figure*}

\begin{figure*}
\begin{flushleft}
  \includegraphics{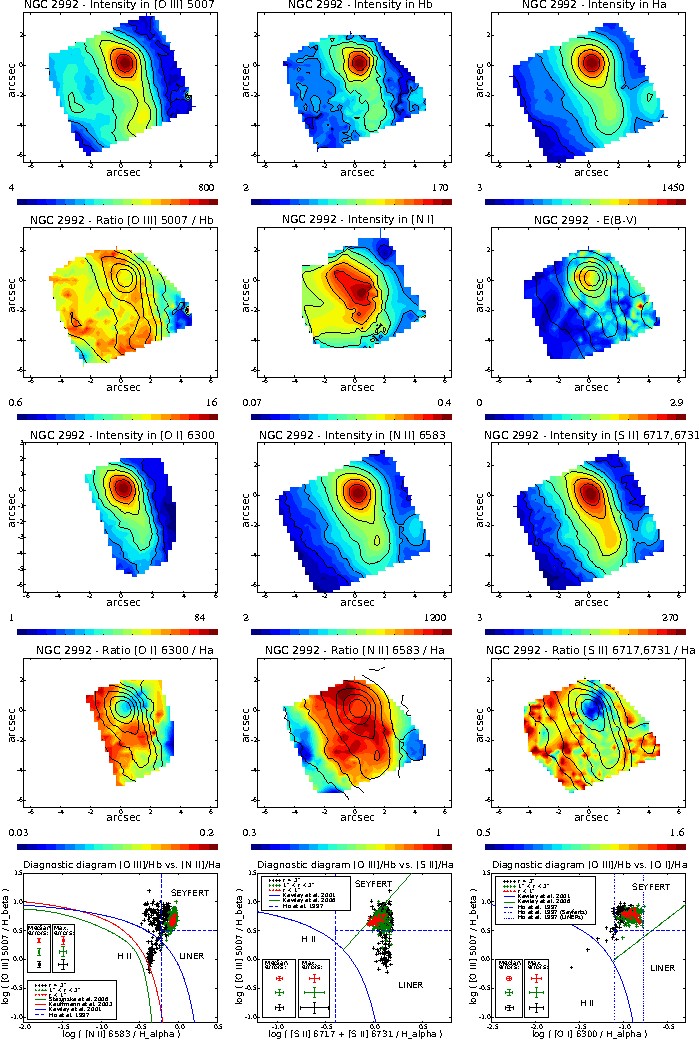}
\caption{\label{NGC2992maps} NGC 2992.
}
\end{flushleft}
\end{figure*}

\begin{figure*}
\begin{flushleft}
  \includegraphics{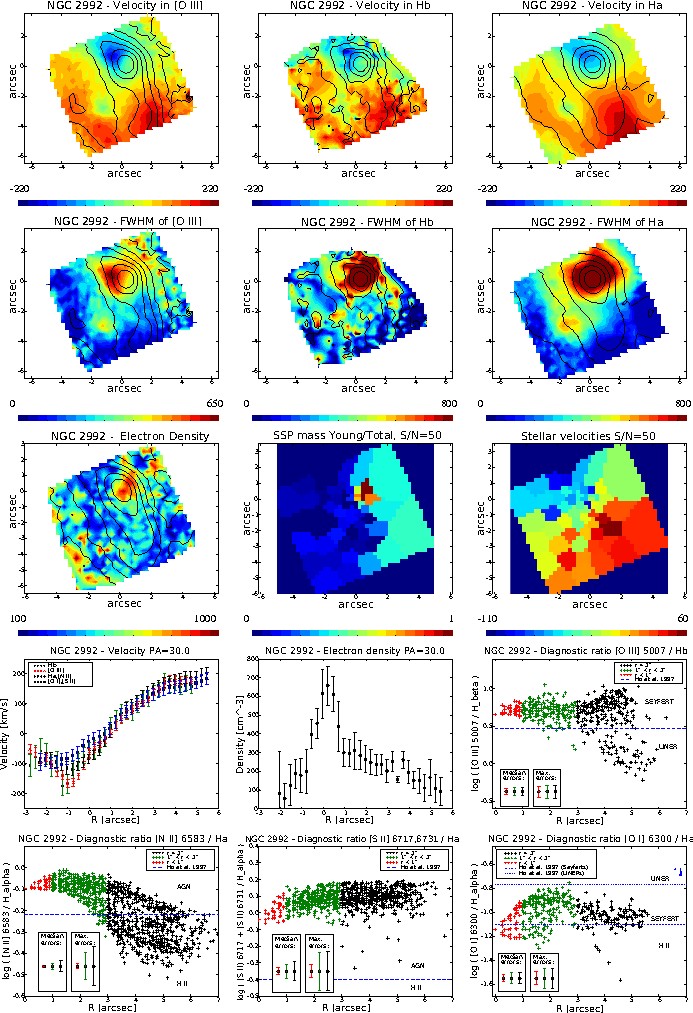}
\caption{\label{NGC2992vel} NGC 2992.
}
\end{flushleft}
\end{figure*}

\begin{figure*}
\begin{flushleft}
  \includegraphics{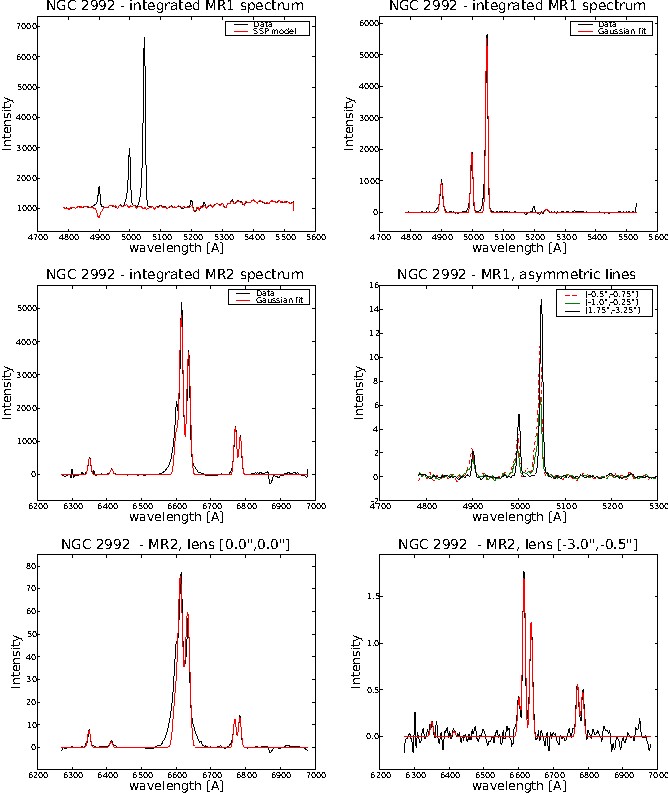}
\caption{\label{NGC2992spec} NGC 2992.
}
\end{flushleft}
\end{figure*}

\begin{figure*}
\begin{flushleft}
  \includegraphics{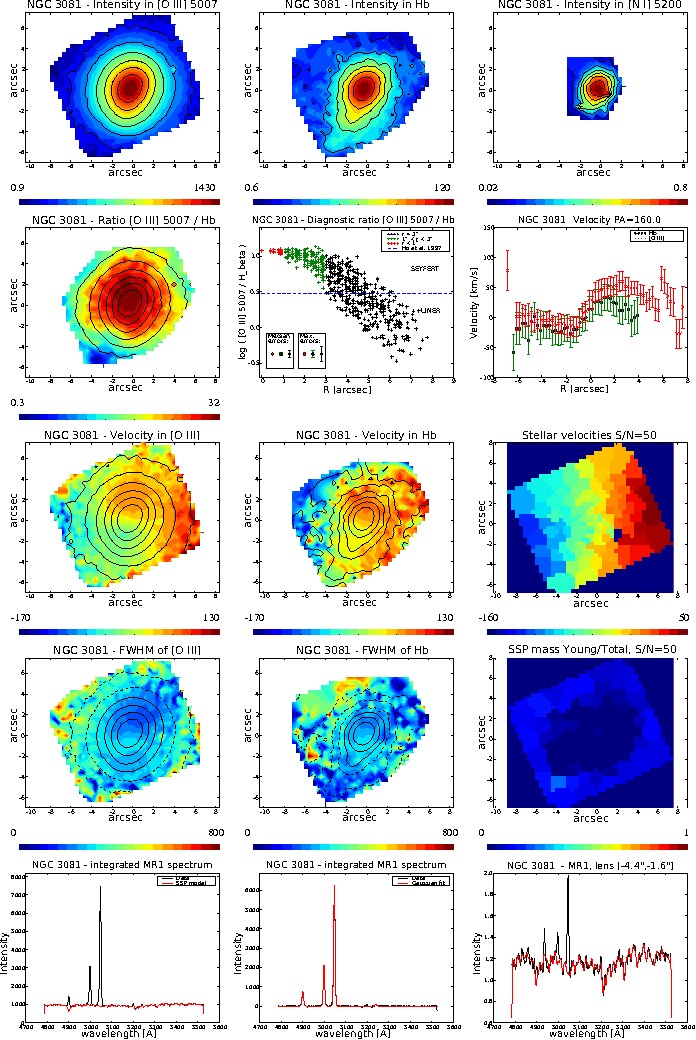}
\caption{\label{NGC3081maps} NGC 3081.
}
\end{flushleft}
\end{figure*}

\begin{figure*}
\begin{flushleft}
  \includegraphics{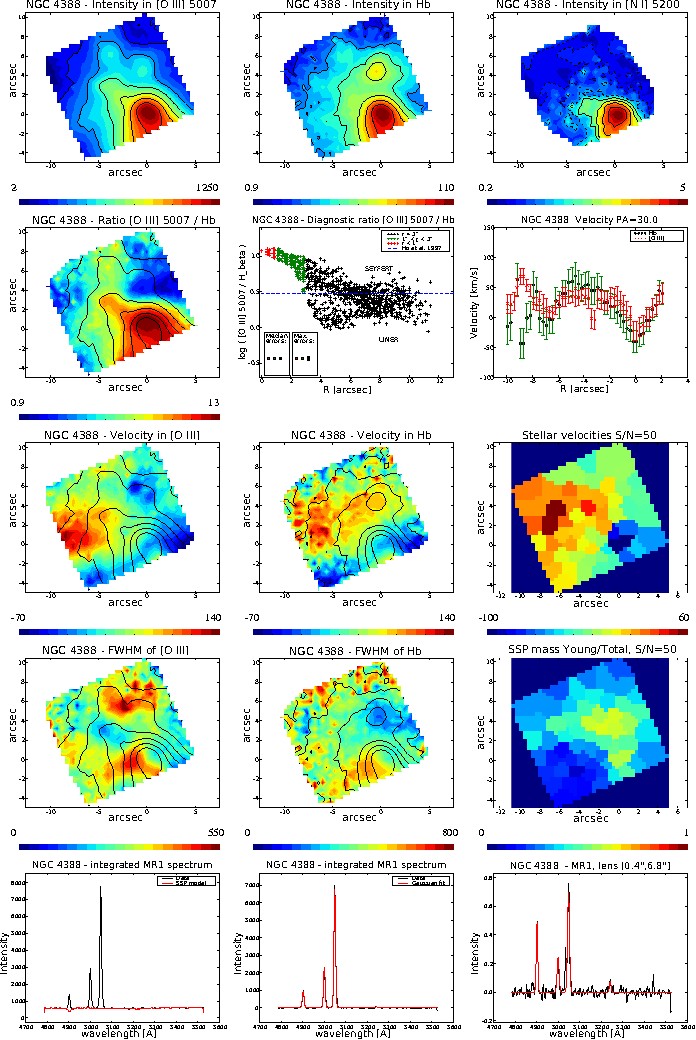}
\caption{\label{NGC4388maps} NGC 4388.
}
\end{flushleft}
\end{figure*}

\begin{figure*}
\begin{flushleft}
  \includegraphics{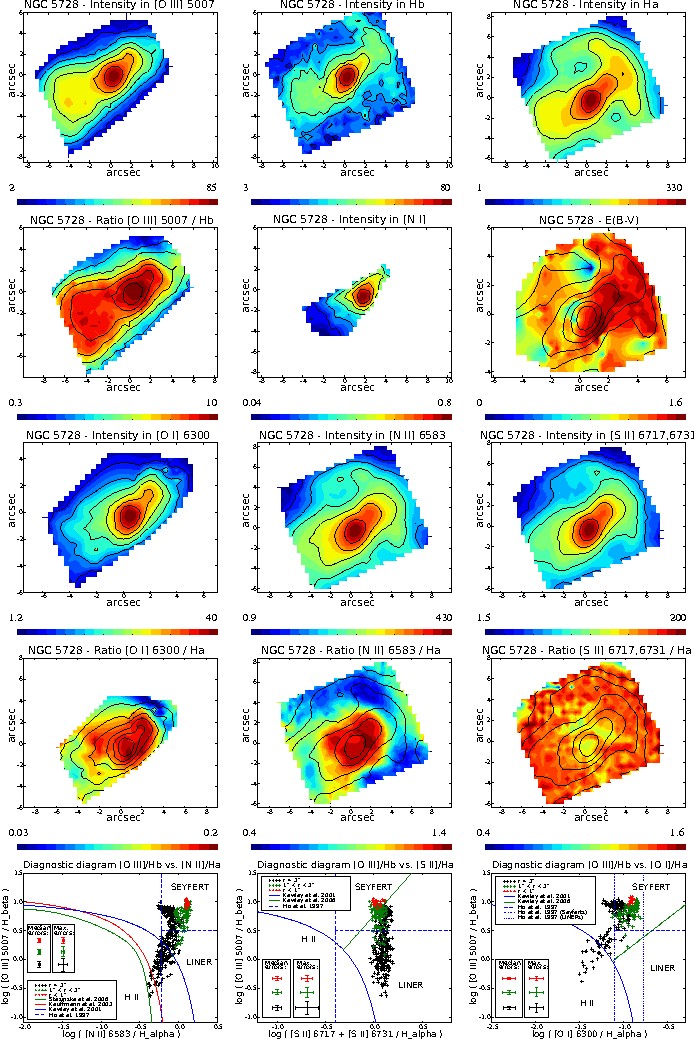}
\caption{\label{NGC5728maps} NGC 5728.
}
\end{flushleft}
\end{figure*}

\begin{figure*}
\begin{flushleft}
  \includegraphics{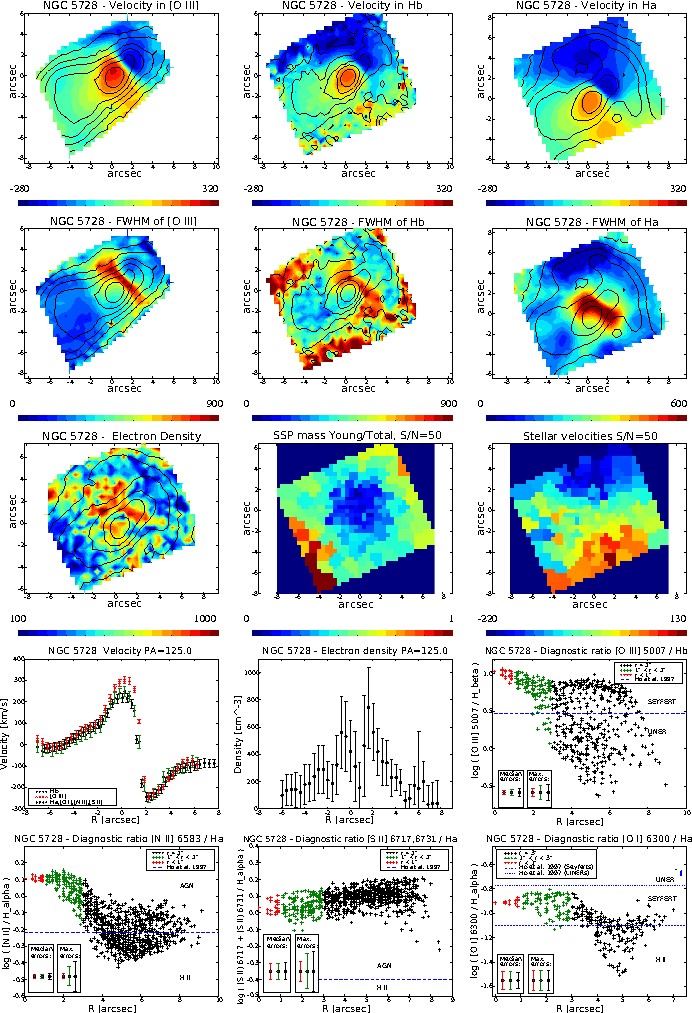}
\caption{\label{NGC5728vel} NGC 5728.
}
\end{flushleft}
\end{figure*}

\begin{figure*}
\begin{flushleft}
  \includegraphics{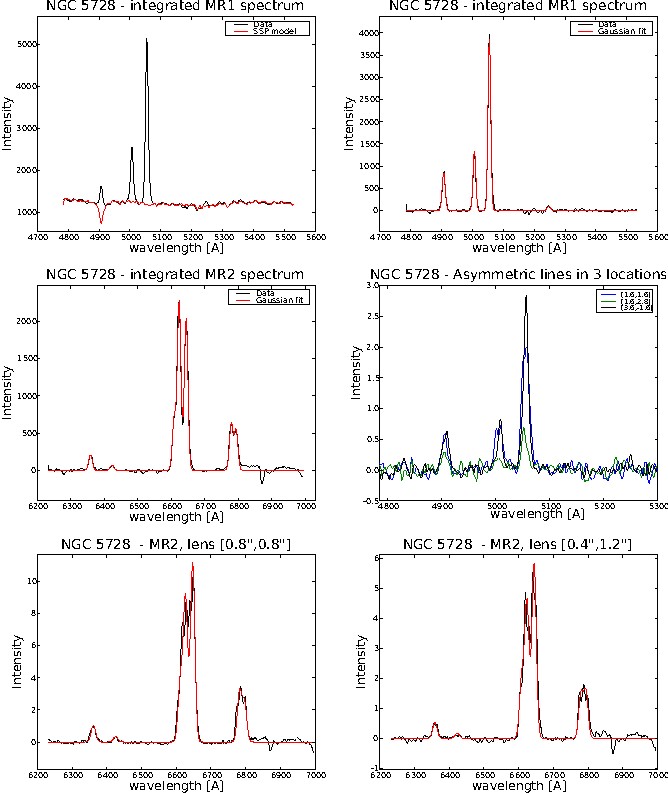}
\caption{\label{NGC5728spec} NGC 5728.
}
\end{flushleft}
\end{figure*}

\begin{figure*}
\begin{flushleft}
  \includegraphics{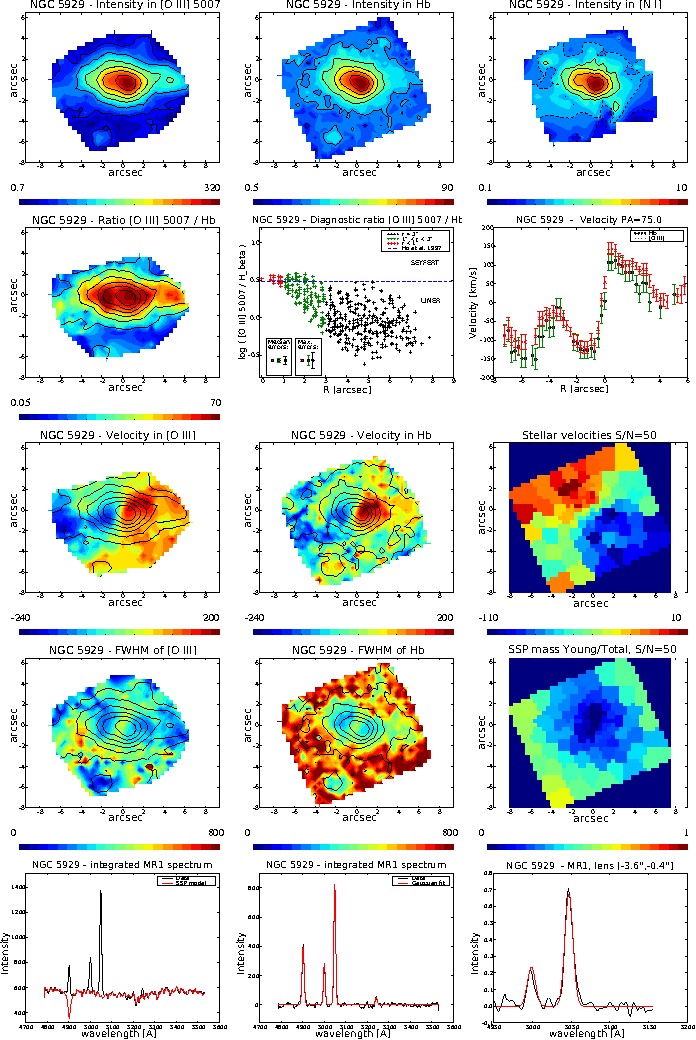}
\caption{\label{NGC5929maps} NGC 5929.
}
\end{flushleft}
\end{figure*}

\noindent
{\bf Caption to Figs.\,\ref{Mrk34maps}-\ref{NGC5929maps} }
\\

Figures \ref{Mrk34maps}-\ref{NGC5929maps} 
contain (where available and relevant to 
the des\-cription of results and discussion, 
depending on individual galaxies):
\\

a) 2D maps of surface brightness in emission lines of
{H$\alpha$}, {H$\beta$}, [\ion{O}{iii}], [\ion{N}{i}], 
[\ion{O}{i}], [\ion{N}{ii}], and [\ion{S}{ii}]
(in units of $10^{-19}$\,J\,s$^{-1}$\,arcsec$^{-2}$\,m$^{-2}$),
mean LOS velocity (in km\,s$^{-1}$), 
velocity dispersion (FWHM, in km\,s$^{-1}$),
line intensity ratios [\ion{O}{iii}]/{H$\beta$}, [\ion{N}{ii}]/{H$\alpha$}, 
[\ion{S}{ii}]/{H$\alpha$}, 
[\ion{O}{i}]/{H$\alpha$},
reddening $E(B-V)$, and electron density;
\\

b) Spatially resolved spectral-diagnostic diagrams;
\\

c) Selected emission-line ratios as functions of projected 
galactocentric distance; 
\\  

d) 1D cuts through mean LOS velocity and electron density along 
specific position angles; 
\\

e) Spectra -- for selected individual spaxels, and/or integrated 
over circular apertures or the entire FOV -- and their fits.   
\\

In the maps, north is up, east to the left. The black contours 
represent the surface brightness in the corresponding emission line for each of
the derived quantities: in the case of mapped line ratios, the contours
correspond to the forbidden line included in the ratio; for the electron 
density, they represent the 
sum of the two sulphur lines; and for the maps of interstellar reddening, 
they depict the surface brightness in {H$\alpha$}. Any discrepancies between
contours of the same species in different maps result from their 
re-computation     
for the new grid, where points of low S/N in the other relevant line(s) were
removed.      

In the 
one-dimensional cuts, the projected radial-distance co-ordinates
have been defined as increasing from east to west. 
Error bars in diagnostic diagrams and the emission-line ratio scatter plots 
have not been assigned to individual points, for clarity. Instead, median and 
maximum 
errors have been plotted for each of the defined spatial regions, in the
corresponding colours.           
The spatially resolved spectral-diagnostic diagrams compare several 
classification schemes commonly applied 
to emission objects (originally with spatially non-resolved spectra). 
The two parallel lines 
for [\ion{O}{i}]/{H$\alpha$} correspond to the values 
provided by \citet{HoIII}, 
$\log$({[\ion{O}{i}]/{H$\alpha$}}) $\geq -1.1 $ for Seyfert galaxies, and  
$\log$({[\ion{O}{i}]/{H$\alpha$}}) $\geq -0.77 $ for LINERs.

\clearpage



\clearpage

\clearpage

\clearpage

\clearpage

\clearpage

\clearpage

\clearpage

\clearpage

\clearpage

\clearpage

\clearpage

\clearpage

\clearpage

\clearpage

\clearpage

\clearpage

\clearpage

\clearpage

\clearpage

\clearpage

\clearpage

\clearpage

\clearpage

\end{document}